\documentclass[12pt, english]{article}
\usepackage[T1]{fontenc}
\usepackage[english]{babel}
\usepackage[letterpaper]{geometry}
\usepackage{fourier}

\usepackage{microtype} 
\usepackage{graphicx}
\usepackage{array}
\usepackage[longnamesfirst,authoryear]{natbib}
\usepackage[textfont=sc,labelfont=bf]{caption}
\usepackage{setspace}
\usepackage{multirow}
\usepackage{amsthm}
\usepackage{amsmath}
    \allowdisplaybreaks[1] 
\usepackage{amsfonts}
\usepackage{amssymb}
\usepackage{svg}
\usepackage{pgf} 
\usepackage{bm}
\usepackage{paralist}
\usepackage{afterpage}
\usepackage{pdfpages}
\usepackage{xfrac}
\usepackage{arydshln}
\usepackage{tabularx}
\usepackage{epstopdf}
\usepackage{comment}

\usepackage{titlesec}
    \titleformat*{\section}{\Large\bfseries}
    \titleformat*{\subsection}{\large\bfseries}
    \titleformat*{\subsubsection}{\large\bfseries}

\usepackage{xcolor}
    \definecolor{darkblue}{RGB}{30,30,130}
    \definecolor{darkred}{RGB}{130,10,10}
    \definecolor{midblue}{RGB}{120,120,255}
    \definecolor{midred}{RGB}{255,120,120}
    \definecolor{lightred}{RGB}{255,227,227}
    \definecolor{amaranth}{RGB}{230,43,79}
    \definecolor{cardinal}{RGB}{196,31,59}
    
    \definecolor{tab_blue}{RGB}{31,119,180}
    \definecolor{tab_orange}{RGB}{255,127,14}
    \definecolor{tab_green}{RGB}{44,160,44}
    \definecolor{tab_red}{RGB}{214,39,40}
    \definecolor{tab_purple}{RGB}{148,103,189}
\usepackage[unicode=true]{hyperref}
    \hypersetup{colorlinks=true,linkcolor=darkred,citecolor=darkred,urlcolor=darkblue}

\newcolumntype{L}[1]{>{\raggedright\let\newline\\\arraybackslash\hspace{0pt}}m{#1}}
\newcolumntype{C}[1]{>{\centering\let\newline\\\arraybackslash\hspace{0pt}}m{#1}}
\newcolumntype{R}[1]{>{\raggedleft\let\newline\\\arraybackslash\hspace{0pt}}m{#1}}
\DeclareMathAlphabet\mathbfcal{OMS}{cmsy}{b}{n}

\newcommand{\beq}{\begin{equation}}
\newcommand{\eeq}{\end{equation}}
\newcommand{\bea}{\begin{eqnarray}}
\newcommand{\eea}{\end{eqnarray}}
\newcommand{\ba}{\begin{array}}
\newcommand{\ea}{\end{array}}
\newcommand{\bit}{\begin{itemize}}
\newcommand{\eit}{\end{itemize}}
\newcommand{\ben}{\begin{enumerate}} 
\newcommand{\een}{\end{enumerate}}
\newcommand{\bpm}{\begin{pmatrix}}
\newcommand{\epm}{\end{pmatrix}}
\newcommand{\bbm}{\begin{bmatrix}}
\newcommand{\ebm}{\end{bmatrix}}

\newtheorem{fact}{Fact}

\makeatletter
\numberwithin{equation}{section}
\theoremstyle{plain}
\newtheorem*{prop*}{\protect\propositionname}
\makeatother
\providecommand{\propositionname}{Proposition}

\providecommand{\tabularnewline}{\\}

\newif\ifreview
\reviewfalse
\ifreview
\paperwidth=\dimexpr \paperwidth + 3cm\relax
\oddsidemargin=\dimexpr\oddsidemargin \relax
\evensidemargin=\dimexpr\evensidemargin + 3cm\relax
\marginparwidth=\dimexpr \marginparwidth + 3cm\relax
\usepackage{todonotes}
\else
\usepackage[disable]{todonotes}
\fi

\usepackage{titling}
\title{\LARGE \textbf{Stock-driven Household Attention}\thanks{First draft: December 7, 2023. We would like to thank Choongryul Yang for the insightful discussions during the early stages of this project. We also appreciate the valuable feedback provided by Pei Kuang, Yeji Sung, and other participants at ASSA 2024. The views expressed herein are those of the authors, and do not reflect the views of the Federal Reserve Board or any person associated with the Federal Reserve System.} }
\author{%
    Hie Joo Ahn\thanks{Federal Reserve Board of Governors, 20th Street and Constitution Avenue NW, Washington, DC 20551, U.S.A. Email: \protect\href{mailto:econ.hjahn@gmail.com}{econ.hjahn@gmail.com}} \\[-0.25cm] 
    \normalsize Federal Reserve Board \and
	Shihan Xie\thanks{Department of Economics, University of Illinois at Urbana-Champaign, 1407 W Gregory Dr, MC-707, Urbana, IL 61801, U.S.A. Email: \protect\href{mailto:shihan.xie@gmail.com}{shihan.xie@gmail.com}} \\[-0.25cm]
	\normalsize University of Illinois \\[0.5cm]
}  
\date{\normalsize \today} 

\begin{document}
    \maketitle
    \thispagestyle{empty}
    \vspace*{-0.8cm}
    \begin{abstract}
    \begin{spacing}{1.2}
    \noindent  
      We investigate the effects of stockholding on households’ attention to the macroeconomy. Households’ attentiveness is measured by their accuracy of inflation expectations and perceptions.  Relative to non-stockholders, stockholders produce more accurate inflation forecasts and backcasts, disagree less about future inflation, and adjust their outlook more responsively to news, suggesting that stock-market participation raises households' attention. Frequent changes in stock prices incentivize stockholders to closely monitor financial markets for optimal trading, given the low cost of acquiring information. Consequently, paying attention to the macroeconomy helps hedge the risks associated with holding stocks. Therefore, attention heterogeneity driven by stockholdings can be a channel through which the distributional consequences of monetary policy are created. 
    \end{spacing}
    \vspace*{0.5cm}
    
    \noindent \textit{JEL classification:} D83, D84, E31, G51 \\[-0.2cm]
    \noindent \textit{Keywords:} Inflation expectations, Household finance, Limited attention
    \end{abstract}

\newpage


\setcounter{page}{1}

\section{Introduction} 

Household heterogeneity is an important factor behind aggregate economic fluctuations and also has profound effects on the transmission of monetary policy (e.g., \citealp{cloyne2020monetary,luetticke2021transmission}). One important source of heterogeneity is household expectations of future macroeconomic conditions. Despite the centrality of household expectations in determining consumption, labor supply, and asset prices, the mechanisms behind these expectation formation processes remain under-explored.

This paper focuses on stock-market participation as an important heterogeneity in household expectation formation. Specifically, we investigate this research question:  \emph{``How does households' stockholding affect economic agents' expectation formation?''} This question is crucial because portfolio choices or stock-market participation, traditionally seen as outcomes of economic agents' risk preferences and ex-ante attentiveness to new information (e.g., \citealp{luo2017robustly,kuhnen2017socioeconomic}), could also serve as determinants of these very factors. We propose that households' stockholding behavior generates an endogenous incentive to monitor macroeconomic news, thus enhancing their attentiveness.

Our main argument is that stockholding creates a significant incentive for households to stay informed about macroeconomic conditions. Stockholders are motivated to monitor economic news closely due to the frequent changes in stock prices and relatively low costs of transactions and information acquisition. Moreover, households can hedge the risk associated with stockholding by paying attention to relevant economic news, which creates an additional incentive for attention. This feedback effect has profound implications for monetary policy, particularly in terms of its effectiveness and distributional consequences.

Using individual-level data from the University of Michigan Survey of Consumers (MSC), we empirically investigate the extent to which stockholding influences households' attentiveness to macroeconomic news. We measure attentiveness with the accuracy of inflation expectations and perceptions of recent past inflation. Additionally, we construct individual-level indicators of attentiveness and attitude towards economic conditions. These indicators allow us to establish new empirical regularities: stockholders' inflation forecasts and backcasts are more accurate, they exhibit less disagreement about inflation, and they adjust their economic outlook more responsively to macroeconomic shocks compared to non-holders.

Our findings suggest that stock-market participation significantly enhances households' attentiveness to macroeconomic news, which in turn improves the accuracy of their economic forecasts. This attentiveness is not merely a byproduct of intrinsic risk preferences or optimism but is an endogenous response to the demands of effective stock market participation. Furthermore, we demonstrate that during periods of heightened economic uncertainty, stockholders intensify their focus on economic news more than non-holders, thereby supporting the risk-hedging motive.

As mechanisms through which stockholdings increase an individual's attention to the macroeconomy, we highlight the \emph{frequency channel} and the \emph{risk-hedging channel}. 
To substantiate what we term \emph{the frequency channel}, we present evidence based on households' gasoline price projections. Notably, both stockholders and nonholders show similar accuracy in their gasoline price projections. Gasoline, being an essential part of households' consumption, directly affects household finances regardless of stock ownership. Given the frequent changes in gas prices, households actively seek the optimal timing of purchasing gasoline, with information on price changes being readily accessible at local gas stations. As a result, both groups pay attention to gasoline price movements and produce equally accurate forecasts. This evidence suggests that high-frequency price changes in critical consumption goods drive households to acquire pertinent information, particularly when the costs of information acquisition and transaction are sufficiently low.

Additionally, \emph{the risk-hedging} motive provides further explanation for our empirical findings. Stocks, as inherently risky assets, are more likely to be held by less risk-averse households. Nonetheless, stockholders can still mitigate the uncertainty or risk associated with stock holdings by raising attention to relevant economic news. In this sense, the cost of attention is analogous to the price of a put option or that of shorting a futures contract in that the acquired information reduces the downside risk of stock holdings. In fact, in times of heightened macroeconomic or personal uncertainty, stockholders raise their attention to economic news more than non-holders do, suggesting that risk-hedging motivation is stronger among stockholders than among non-holders.  

These results highlight an important channel through which monetary policy can have differential effects on stockholders and non-holders. Stockholders, being more informed and attentive, are likely to adjust and re-optimize their consumption and investment decisions more promptly in response to monetary policy changes. This dynamics potentially amplifies the wealth inequality between stockholders and non-holders, raising important considerations for policymakers. Our empirical evidence supports this by showing the dynamic responses of the two groups' economic outlooks to monetary policy shocks. For instance, stockholders' expectations about inflation and the real economy respond more significantly and immediately to monetary policy shocks compared to non-holders. These findings underscore the necessity for policymakers to consider the distributional effects of monetary policy and the role of household asset portfolios in shaping these effects.


Our paper contributes to the literature on expectation heterogeneity and its effects on household behavior (e.g., \citealt{itzhak2018}).
A recent study by \cite{LiSinha2023} finds that the sentiments of homeowners and stockholders are more sensitive to expected inflation than those of other consumers, revealing a disparity that challenges the notion that owning such assets provides hedges against inflation.
While macroeconomic literature has examined household heterogeneity, including socio-economic status, homeownership, and wealth in expectation formation (e.g., \citealp{das2019, Mitman2021, Xie2023, AXY}), no prior studies have considered the impact of stock-market participation on attentiveness as a channel through which monetary policy has distributional effects on household welfare. Our research fills this gap by demonstrating that stockholders are more attentive and produce more accurate economic forecasts, thereby amplifying the distributional effects of monetary policy.

Our research extends the growing literature on household attention to macroeconomic conditions (e.g. \citealt{link2024attention,BrachaTang2022}). This paper is the first one that focuses on stock-market participation as a driver of household attention heterogeneity. 
Relatedly, \cite{AXY} investigates the role of homeownership in households' attention and finds that homeowners are better informed about monetary policy changes than renters due to mortgage-holding. We provide evidence that stockholders produce more accurate inflation projections than homeowners, suggesting that stock-market participation likely provides a greater incentive to pay attention to more general economic news compared to homeownership.

Finally, our paper makes unique contributions to the macro-finance literature by providing novel insights into theories about expectation formation. We highlight a new perspective where stock ownership --- traditionally viewed as the outcome of households' optimization (e.g., \citealp{luo2010rational,luo2017robustly}) --- can also be an important determinant of households' attention. Stockholding induces feedback effects on household expectation formation (e.g., \citealp{kuhnen2015asymmetric}), amplifying the distributional effects of monetary policy. A theoretical implication is that stockholding can be a key channel through which attention to economic news is endogenously determined, making it a crucial source of heterogeneity in rational inattention models. This stock-driven attention channel, not previously explored in the literature, opens new avenues for future research.
Our research distinguishes itself from the finance literature that focuses on the role of expectations heterogeneity in portfolio choices (e.g., \citealp{kuhnen2017socioeconomic}). For example, \cite{huang2007rational} argue that economic agents' portfolio choices are suboptimal owing to the information acquisition costs. In contrast, we focus on reverse causation: the consequence of stock market participation on household attentiveness and macroeconomic expectations. 


This paper is structured as follows. 
Section \ref{sec:data} discusses the data for empirical analyses. 
Section \ref{sec:empirics} presents empirical evidence on the relationship between stock market participation and household attention to macroeconomic conditions. 
Section \ref{sec:mechanism} discusses the mechanism behind the empirical evidence.  
Section \ref{sec:implications} discusses implications for theory and policy. 
Section \ref{sec:conclusion} concludes.

\section{Data}\label{sec:data} 

This section discusses data for our empirical analyses. Section \ref{ss:ie_data} discusses individual-level expectations data from the Michigan Survey of Consumers (MSC). Section \ref{ss:attn} introduces the individual-level measure of attentiveness and attitude.  

\subsection{Michigan Survey of Consumers} \label{ss:ie_data} 

The individual-level data from the Michigan Survey of Consumers (MSC) is available monthly as repeated cross-sectional data. The dataset contains a small subset with panel dimension as 40\% of the respondents are re-contacted once after 6 months. The dataset comes along with a rich set of individual demographics. Our main analysis uses MSC microdata for the sample of 1997:M1 -- 2021:M12. We discuss a few key features of the data for our study.

\paragraph{Stock-market participation} Information about stock-market participation has been available since 1997. The survey uses this following question to measure stock market participation: 
\begin{quote}
    ``The next questions are about investments in the stock market.
First, do you (or any member of your family living there) have
any investments in the stock market, including any publicly
traded stock that is directly owned, stocks in mutual funds,
stocks in any of your retirement accounts, including
401(K)s, IRAs, or Keogh accounts?''
\end{quote}
Those who say \emph{``Yes''} to the question are classified as stock-market participants (henceforth \textit{stockholders}) and the rest are classified as nonparticipants (henceforth \textit{non-holders}). On average, about 60\% of the respondents in MSC are stockholders.
This dataset, however, does not provide information about whether an individual recently traded stocks or financial products that include stocks. 

\paragraph{Inflation forecast error} We are primarily interested in the accuracy of household inflation forecasts for a few reasons. First, the forecast errors reflect the degree of individual attentiveness and also the outcome of attentiveness. Second, numeric responses are recorded to questions about inflation expectations, and hence the accuracy of forecasts is quantitatively evaluated unlike other expectations expressed qualitatively. We measure the accuracy by comparing their reported expectations 12 months ahead against the corresponding realized CPI inflation rate.

\paragraph{Perception of current inflation} 
We also use the special supplement to MSC that asks about households' perception of the inflation level in the recent past (the \emph{backcast}, henceforth).\footnote{The Board of Governors of the Federal Reserve System initiated and sponsored new questions on inflation perceptions in 2016. The questions on perceptions are worded consistently with the questions on inflation expectations and are currently posed four times a year--in February, May, August, and November. This dataset is not publicly available. See \cite{Axelrod2018} for the data description.} This supplement is quarterly, and the sample period is 2016:Q1 -- 2021:Q4, shorter than the sample period of the baseline MSC dataset. 
\begin{quote}
    ``During the past 12 months, do you think that prices in general 
went up or went down, or stayed where they were a year ago? By about what percent do you think prices went (up/down), on average, during the past 12 months?''
\end{quote}
We also measure the accuracy of backcasts by comparing them against the corresponding realized CPI inflation rate. As suggested in \cite{BrachaTang2022}, inflation backcast error is a better measure of individual attention to inflation than forecast error since it is not driven by the effects of an inability to form a forecast of future inflation.


\subsection{Measures of attentiveness and attitudes} \label{ss:attn} 

To facilitate our analysis, we construct novel individual-level indicators that directly measure the effort of information acquisition and the perception of acquired information. Our indicators are based on an under-explored question in the MSC about whether a person recalls any news related to business conditions in the past few months, and whether this person recalls positive or negative news.\footnote{A few existing studies that explored this questions includes \cite{das2019,drager2017imperfect}.} We use the household's response on whether they recall any news to construct the indicator of \textit{attentiveness}, and we use their recall of either positive or negative news to construct the measure of \emph{attitude}. This question has been asked monthly in the survey since 1978M1. Specifically, the following question has been asked: 
\begin{quote}
    ``During the last few months, have you heard of any favorable or unfavorable changes in business conditions? What did you hear?''
\end{quote}
We create a new variable called \emph{attentiveness}, a dummy variable that equals $1$ if households recall at least one change and $0$ if not. 
During the sample period, about 60 percent of households recall at least one change. Among the respondents who recall at least one change, about 18 percent recall only favorable changes, and 34 percent recall only unfavorable changes.\footnote{A small fraction of attentive households recall both good and bad news at the same time.} Additionally, we create another variable called \emph{attitude}, a categorical variable that equals $1$ if households recall only favorable changes, $-1$ if only unfavorable changes, and $0$ otherwise. In \ref{app:index}, we show that at the aggregate level, attentiveness is countercyclical whereas attitude is procyclical.\footnote{Countercyclical attention is also consistent with findings in \cite{coibion2015information}. Relatedly, \cite{BrachaTang2022} find consumers pay greater attention when inflation is high based on individual-level responses to inflation-related questions from MSC.}

\paragraph{Attentiveness, attitude, and inflation forecasts}
We validate our individual-level attentiveness and attitude measures by documenting several facts.\footnote{\ref{app:att} provides more details on the exercises.} We find an upward bias in households' inflation expectations, consistent with existing literature (\citealp{upbias_jep}). On average, attentive households report lower and more accurate inflation forecasts. Column (1) in Table \ref{tab:attitude_fe} reports the estimation result.  
Next, we investigate the properties of our measures of attentiveness and attitudes with other related individual-level indices developed using MSC. Column (2) of Table \ref{tab:attitude_fe} shows that more attentive individuals are also more certain about their inflation expectations using the inflation uncertainty measure developed in \cite{binder2017measuring}. 
Column (3) of Table \ref{tab:attitude_fe} shows that individuals who are in general attentive to macroeconomic conditions are less likely to be inattentive to current inflation using the inflation inattention measure developed in \cite{BrachaTang2022}.

\section{Empirical Evidence} \label{sec:empirics} 

\subsection{Stylized Facts} \label{sec:stylized}
We establish several stylized facts regarding households' stock-market participation and its effect on their attentiveness to macroeconomic conditions. First, we evaluate the accuracy of inflation forecasts and backcasts. Second, we study disagreement about inflation expectations within different household groups. Third, we focus on the responsiveness of macro expectations to various economic shocks. 


We first examine the forecast accuracy of four groups of households: those who own stocks and a home (homeowner and stockholder); those who own stocks and do not own a home (renter and stockholder); those who do not own stocks and own a home (homeowner and non-holder); those who do neither own stocks nor a home (renter and non-holder). The top panel of Figure \ref{fig:error} displays average absolute errors in forecasting 12-month-ahead inflation of the four groups. 
Notably,  forecast errors of stock market participants are consistently smaller than those of non-participants regardless of the homeownership status. 
The difference in accuracy between stockholders and non-holders is larger than that between homeowners and renters, as shown in the middle and bottom panels of Figure \ref{fig:error}. Column (1) of Table \ref{tab:stylized} provides the average forecast error of each household group over the entire sample period.

To further trace the source of this difference, we look into each of the four groups' accuracy of inflation backcasts. 
Figure \ref{fig:error_backcasts} reports the backcast errors. 
Like inflation forecasts, stockholding is the key factor that reduces backcast errors. 
Notably, households who own both stocks and a home produce the most accurate backcasts, and those who own stocks but do not own a home make comparable errors. 
Households who own neither stock nor home make the largest backcast errors, followed by households who own a home but not stocks. 
Backcast errors of those who only own houses are about twice as large as those of individuals who only own stocks. 
This observation suggests that stockholding improves the perception of inflation in the recent past, which eventually translates into the accuracy of inflation forecasts. Homeownership raises households' attention to the macro economy, but its effect is not as strong as that of stock market participation. 

\begin{fact}
    Stockholders' expectations of future inflation and perceptions of recent past inflation are more accurate than those of non-holders.
\end{fact}


Next, we examine disagreements about inflation forecasts and backcasts within each group. Disagreement is informative about heterogeneity in attentiveness and in the accuracy of information that households have about overall economic conditions. In the literature on expectation formation, disagreement is an empirical test bed where various macroeconomic theories on information friction are tested (\citealp{CG2012JPE}). 

The within-group disagreement is measured by each group's standard deviation of inflation forecasts and that of backcasts. The top of Figure \ref{fig:sd} displays the within-group disagreement by asset-holding status. Stockholders disagree less about future inflation expectations than nonholders regardless of homeownership. Differences in disagreement between stockholders and non-holders are larger than those between homeowners and renters, as shown in the middle and bottom panels of Figure \ref{fig:sd}. The substantially larger disagreement among non-holders also suggests that macroeconomic news has a weak influence on household expectation formations among them. Columns (2) and (3) of Table \ref{tab:stylized} provide the average standard deviation and inter-quartile range of each household group over the entire sample period. 

Similar patterns are observed in the backcasts, displayed in Figure \ref{fig:sd_backcasts}. Those who own both stocks and homes disagree the least, though those who own stocks but do not own a home exhibit comparable within-group disagreement and even disagree less at times. Meanwhile, households who do not own stocks show much larger within-group disagreement, whereas the group who own neither stocks nor a home shows the largest within-group disagreement. All told stock-market participation reduces the disagreement about inflation backcasts, suggesting that stockholding incentivizes households to be better informed of the macroeconomic conditions than in the case of no stockholding.\footnote{Note that a home is an important asset in households' finances, and homeowners respond to interest-rate changes more than renters (\citealp{AXY}). Notably, stockholding creates significantly larger differences in the accuracy of inflation projection and greater discrepancies in the degree of disagreement than homeownership does.} 

\begin{fact}
    Each group shows sizable disagreement about future inflation and recent-past inflation, but the within-group disagreement is larger among non-holders than among stockholders.
\end{fact}

\begin{table}
    \centering
    \caption{Summary of stylized facts} \label{tab:stylized}
    \small
    \centering
    \vspace{0.3cm}
	\noindent\resizebox{0.95\textwidth}{!}{
	\begin{tabularx}{\textwidth}{l>{\centering}p{3.5cm}>{\centering}p{3.5cm}>{\centering}p{3.5cm}}
        \hline\hline
        \noalign{\vskip0.1cm}
           & (1) Forecast error & (2) Std. dev. & (3) IQR \tabularnewline
        \noalign{\vskip0.1cm}
        \hline 
        \noalign{\vskip0.1cm}
        Homeowner \& Stock holder & 2.77 & 3.46 & 3.49 \tabularnewline
        Renter \& Stock holder  & 2.91 & 3.72 & 3.62 \tabularnewline
        Homeowner \& Nonparticipant & 3.42 & 4.24 & 4.24 \tabularnewline
        Renter \& Nonparticipant & 3.65 & 4.49 & 4.64 \tabularnewline
        \hline 
        \noalign{\vskip0.1cm}
        Sample period        & \multicolumn{3}{c}{1997Q3 - 2021Q2}   \tabularnewline
        \noalign{\vskip0.1cm}
        \hline \hline 
    	\noalign{\vskip0.1cm}
    	\multicolumn{4}{p{\dimexpr\linewidth-2\tabcolsep-2\arrayrulewidth}}{\footnotesize \textit{Notes:} This table provides summary statistics on the forecast accuracy and within-group disagreement over different household groups. Columns (1) - (3) present the time-series mean of inflation forecast error, standard deviation, and inter-quartile range of inflation forecasts.}    \tabularnewline
    	{\footnotesize \textit{Sources:} Authors' calculation.} 
	\end{tabularx}
    }
\end{table}


Lastly, we examine how responsive stockholders and non-holders are to forward guidance, which reflects the outcome of attentiveness.  The effectiveness of forward guidance critically hinges on how well-informed and attentive households are to the direction of future monetary policy, unlike changes in federal funds rates or the implementation of large-scale asset purchases. Employing an event-study approach, we look closely at changes in households' interest-rate predictions after the prominent forward guidance implemented between 2008 and 2019. 
We consider the following four events of forward guidance:
\begin{itemize} \itemsep0em
    \item[\textbf{2008Q4}] The US economy reached the zero lower bound (ZLB) and the FOMC conducted ``open-ended'' forward guidance.
    \item[\textbf{2011Q3}] The FOMC announced that it expected to keep the federal funds rate low at least through mid-2013.\footnote{The professional forecasters' interest rate expectation also began to converge very quickly at this point \citep{andrade2019forward}.}
    \item[\textbf{2013Q2}] The end of the forward guidance period after the Great Recession.
    \item[\textbf{2019Q2}] The FOMC announced that it would steady the policy rate. 
\end{itemize} 


\begin{figure}[ht!]
    \centering
    \caption{Reactions of households to forward guidance}
    \includegraphics[width = 0.5\textwidth]{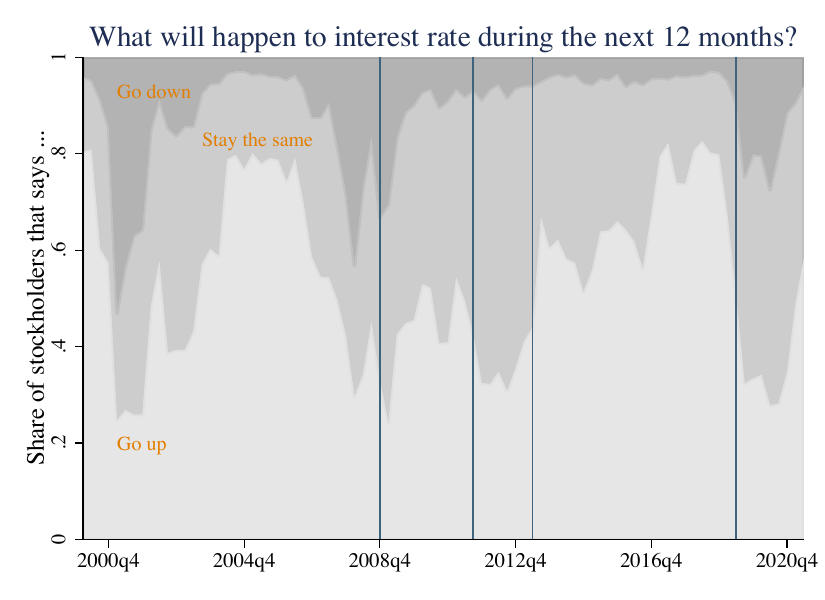}
    \includegraphics[width = 0.5\textwidth]{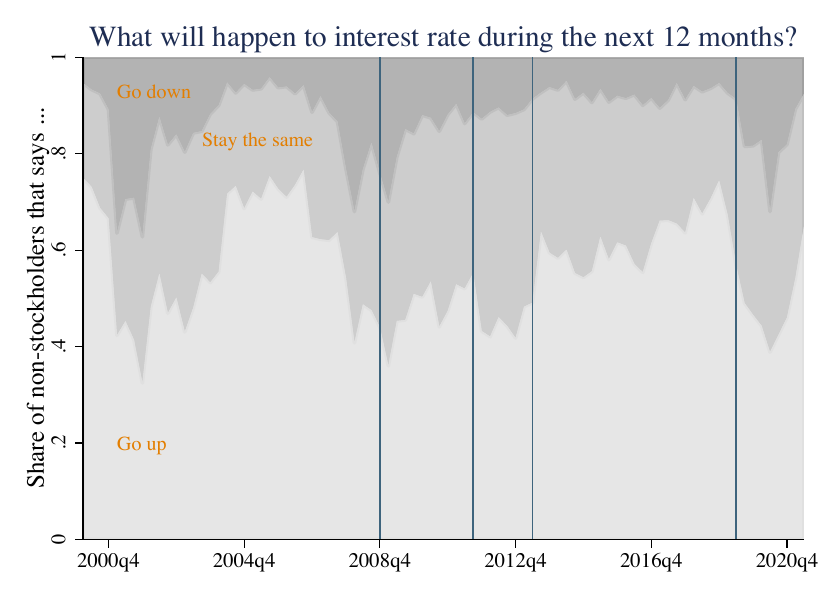}
    \includegraphics[width = 0.5\textwidth]{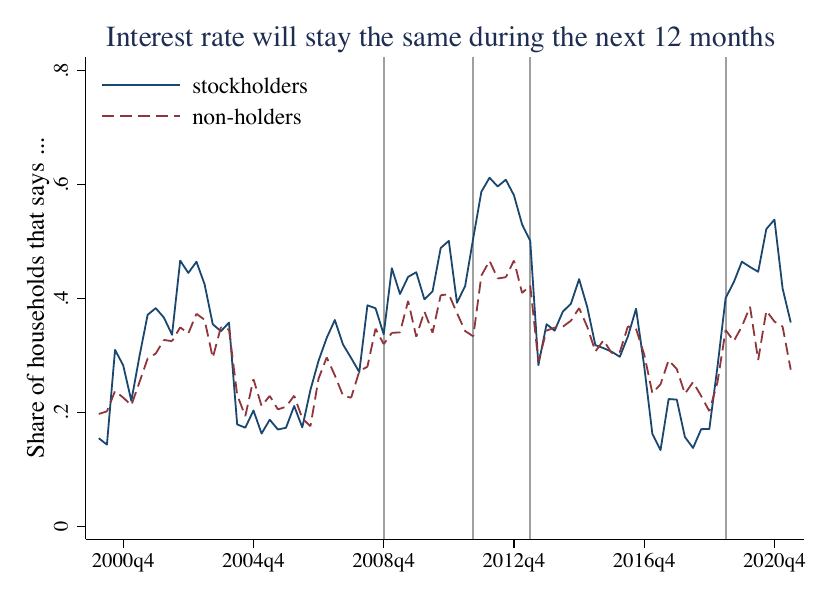}
    \label{fig:forwardguidance}

    \noindent\resizebox{1.0\textwidth}{!}{
		\begin{tabularx}{\textwidth}{m{3.8cm}}
			\noalign{\vskip0.3cm}
			\multicolumn{1}{p{\dimexpr\linewidth-2\tabcolsep-2\arrayrulewidth}}{
				\footnotesize \textit{Notes:} This figure reports the share of households who expect the interest rate would (1) go down; (2) stay the same; and (3) go up during the next 12 months. The four vertical lines from left to right indicate the four events of forward guidance discussed in Section \ref{sec:stylized}. The top panel displays the responses of stockholders and the middle panel shows those of non-holders. The bottom panel displays the shares of stockholders and non-holders who believe that the interest rate will not change in the next 12 months.} \\
			\multicolumn{1}{p{\dimexpr\linewidth-2\tabcolsep-2\arrayrulewidth}}{	
				\footnotesize \textit{Source:} Authors' calculation. 
            }
            \end{tabularx}
    }
\end{figure}


Figure \ref{fig:forwardguidance} reports the share of households who expect the interest rate would (1) go down; (2) stay the same; and (3) go up during the next 12 months. The four vertical lines from left to right indicate the four events of forward guidance. The top panel displays the responses of stockholders and the middle panel shows those of non-holders. The two panels show that stockholders adjust their beliefs more in a way that is consistent with the forward guidance than non-holders do. The bottom panel displays the shares of stockholders and non-holders who believe that the interest rate will not change in the next 12 months. A significantly larger portion of stock-market participants believed that interest rates would stay the same during the next 12 months when the FOMC made those announcements (indicated by the vertical lines). 
In short, stockholders pay more attention to the forward guidance than non-holders, and hence are better informed about future monetary-policy directions. 

\begin{fact}
    Stockholders adjust their economic outlook to macroeconomic events in a more responsive manner than non-holders do.
\end{fact}

\subsection{Regression Analyses}\label{sec:regression} 
 
Next, we present formal statistical analyses in support of our main claim. We focus on differences between stockholders and non-holders to show that stock-market participation on average raises households' attention to macroeconomic news.


We begin with a regression analysis to examine whether stock market participation improves households' overall attentiveness to macroeconomic news. We adopt two measures. First, we examine heterogeneities in inflation forecasts, as the sample period is long enough for reliable statistical analyses, unlike the short sample of backcasts. In addition, the similarity between inflation backcasts and forecasts (\citealp{Axelrod2018}) ensures that the conclusion based on inflation forecasts will likely be similar to that based on inflation backcasts. 
Second, we employ the indicator of \textit{attentiveness} introduced in Section \ref{ss:attn}.

Our regression model is specified as follows: 
\begin{equation}
    Y_{i,t} = \alpha_t + \beta ~\text{\emph{Stock}}_{i,t} + \delta X_{i,t} + \epsilon_{i,t},
\label{eqn:error}
\end{equation}
where the dependent variable $Y_{i,t}$ is respondent $i$'s forecast error about 12-month ahead inflation rates at time $t$ or \textit{attentiveness}, and $\text{\emph{Stock}}_{i,t}$ is a dummy variable that equals $1$ if respondent $i$ participates in stock market and $0$ if not. The notation $X_{i,t}$ captures demographic fixed effects, including gender, education, birth cohort, marriage status, region, homeownership, and income quintiles. Time-fixed effects, denoted by $\alpha_t$, absorb the effect of any variable that varies with macroeconomic conditions but does not vary by individual.
 
\begin{table}
    \centering
    \caption{Stock holding and attentiveness} \label{tab:error}
    \small
    \centering
    \vspace{0.3cm}
	\noindent\resizebox{1\textwidth}{!}{
	\begin{tabularx}{\textwidth}{ll>{\centering}p{1.8cm}>{\centering}p{1.8cm}>
    {\centering}p{1.8cm}>{\centering}p{1.8cm}>{\centering}p{1.8cm}>{\centering}p{1.8cm}}
        \hline\hline
        \noalign{\vskip0.1cm}
           & & \multicolumn{3}{c}{\textbf{Forecast error}} & \multicolumn{3}{c}
         {\textbf{Attentiveness}}  \tabularnewline
        &  & (1) OLS & (2) IV & (3) FD & (4) OLS & (5) IV & (6) FD 
        \tabularnewline[0.1cm]
        \hline 
    	\noalign{\vskip0.1cm}
        Stock ($\beta$)  &  & -0.371$^{***}$ & -0.595$^{***}$ & 0.033 &   0.08$^{***}$ &  0.10$^{***}$ & 0.02$^{***}$ \tabularnewline
                    &   &    (0.024) & (0.061)  & (0.045)
                    & (0.00)    & (0.01) & (0.01)\tabularnewline[0.1cm]
                    \hline 
        \noalign{\vskip0.1cm}
        Demographic FE          & & Y & Y & N & Y & Y  & N
        \tabularnewline[0.1cm]
        Time FE  & & Y & Y & Y & Y & Y & Y
        \tabularnewline[0.1cm]
        No. of obs.  &  &      109,001   & 42,404 &  45,459
        & 120,147  &  45,987 & 54,425 \tabularnewline[0.1cm]
        $R^2$           &   &    0.1157   & 0.0011 & 0.0754 
        & 0.1045   &  0.0038 &  0.0192\tabularnewline[0.1cm]
        \hline\hline
        
    	\noalign{\vskip0.1cm}
    	\multicolumn{8}{p{\dimexpr\linewidth-2\tabcolsep-2\arrayrulewidth}}{\footnotesize \textit{Notes:} This table reports the regression results from Equations (\ref{eqn:error}). Dependent variables are inflation forecast errors (Columns (1)-(3)) and attentiveness (Columns (3)-(5)) respectively. ``Stock'' indicates dummies for stock market participation. Columns (1) and (4) report estimation with OLS. Columns (2) and (5) report 2SLS estimation using stock participation status from six months ago as IV. Columns (3) and (6) report estimation with first-difference (FD). We control for time-fixed effects in all specifications and survey respondents' demographic fixed effects, including gender, education, birth cohort, marriage status, region, homeownership, and income quintiles in OLS and IV regressions. Robust standard errors are reported in the parenthesis.  $^{***}$, $^{**}$, $^{*}$ denotes statistical significance at 1\%, 5\%, and 10\% levels respectively.}    \\ 
    	\multicolumn{8}{p{\dimexpr\linewidth-2\tabcolsep-2\arrayrulewidth}}{\footnotesize \textit{Sources:} Authors' calculation.} 
	\end{tabularx}
    }
\end{table}

Results are reported in Columns (1) and (4) of Table \ref{tab:error}. In Column (1), the coefficient on stock ownership ($\beta$) is negative and statistically significant, indicating that stock-market participation does improve the accuracy of households' inflation forecasts.\footnote{We reach a similar conclusion with households' inflation backcasts.} In Column (2), the coefficient ($\beta$) is positive and statistically significant, suggesting that stock-market participation does raise the attentiveness of an individual to economic news.

\paragraph{Tackling endogeneity issues}
In our baseline regression model, we have controlled a rich set of demographics, as well as time-fixed effects, that correlate with stock ownership status. However, one may still be concerned about potential endogeneity issues due to observed characteristics. 
Namely, an individual's particular unobserved attributes make the person more likely to participate in the stock market and at the same time produce better macroeconomic projections. 
We employ several approaches to tackling the potential endogeneity problem. 

One prominent candidate of the unobserved attribute is ``optimism''. 
Specifically, an individual with a more positive economic outlook may believe that it is a good time to invest in stocks. As a result, this individual starts paying more attention to economic conditions, resulting in more accurate inflation forecasts.
If this optimism about the economic outlook is the key driver of empirical results in Table \ref{tab:error}, it is misleading to claim that stock-market participation motivates households to pay more attention to macroeconomic news and produce more accurate forecasts. To examine the endogeneity concern, we consider the following as robustness checks. 

First, we exploit the sub-sample of households who are re-contacted and use their stock-holding positions from six months ago as an instrument for current stock holding. 
Households' stock-holding decisions from six months ago are less likely to be impacted by the news they heard in recent months. 
The result is still robust, as shown in Columns (2) and (5) of Table \ref{tab:error}, indicating that the endogeneity is less of a concern. 

Second, we include measures of individual attitudes in the regression to directly control the effect of households' optimism about economic conditions. We use two measures to capture individual optimism. One is the new indicator of \textit{attitude} we constructed in Section \ref{ss:attn}. Another one is the sentiment index constructed by MSC.
The results are robust and little changed regardless as shown in Table \ref{tab:attention_control}. 

Third, to further control individual-level unobserved heterogeneity, we account for individual fixed effects in the regression. To do so, we use the first-difference estimator for Equation (\ref{eqn:error}) focusing on the 40\% re-contacted sample. Regression results are reported in Columns (3) and (6) of Table \ref{tab:error}.\footnote{We also consider the New York Fed's Survey of Consumer Expectations (SCE) given that it has a longer panel dimension. However, the SCE does not have information on stock market participation, and therefore, is not suitable for our research question.} One issue with this estimation is that only a small fraction of individuals (less than 7\% in the re-contacted sample) changed their stock market participation status over the 6 months. As a result, there is little variation in the regressor of interests when using the first-difference estimator, giving less accurate estimates of $\beta$. Columns (3) and (6) of Table \ref{tab:error} report the estimation results. The coefficient in Column (3) is highly insignificant due to the issues discussed. The estimation results may also be masked with panel conditioning effects which are prevalent in inflation expectations \citep{kim2023learning}. Nonetheless, the coefficient in Column (6) remains significant and robust, strongly supporting our claim. 
In sum, our evidence suggests that stock-market participation improves the accuracy of inflation projections by raising households' attentiveness to macroeconomic news.

\section{Mechanisms}\label{sec:mechanism} 

We have demonstrated that stock-market participation raises the attentiveness to macroeconomic news. 
In this section, we discuss three channels that contribute to this finding. 
First, the high-frequency information flows make households update their information on the macroeconomy more frequently -- the frequency channel. It is important to note that this channel functions due to the relatively low costs associated with stock trading and information acquisition. These low costs make it beneficial for households to pay attention to macroeconomic news. Second, households' attention to the macroeconomy and the consequent improvement in information precision helps them to hedge against the risk associated with stock holding -- the risk hedging channel.


\subsection{The frequency channel}

All else equal, households have stronger incentives to monitor asset value changes to detect the optimal timing of trading if the value of an asset in their portfolio changes more frequently. 
Stocks are standardized assets, and the majority are traded at open exchanges. The realized prices of stocks change every day and tick by tick. 
Quite differently, homes are heterogeneous, and hence the prices traded at housing markets are not standardized. In addition, home purchases usually involve a long-term mortgage contract. Because of these factors, among other reasons, housing assets are not traded as frequently as stocks, and thus stock prices do not change at a high frequency. 
Therefore, households who are interested in trading stocks or maximizing their wealth via stock trading have more incentive to pay attention to economic news to find the optimal time of trading than those who own a home only as their asset. Consistent with this theory, inflation forecasts of stockholders who do not own a home are more accurate than those of homeowners who do not own stocks (Figures \ref{fig:error} and \ref{fig:error_backcasts}).\footnote{Direct and indirect stock-market participation would also have different effects on households' attentiveness to macroeconomic news. However, the MSC does not have such information.}   

It is important to note that this channel functions effectively only if the costs of acquiring information or trading stocks are sufficiently low. If these costs increase, the net gains from paying attention to economic news and trading stocks would decrease, reducing the incentive to monitor the macroeconomy. In this context, the transaction costs of trading stocks are low, particularly relative to those of purchasing or selling a home. This difference explains why stockholders pay more attention to macroeconomic news and produce more accurate forecasts and backcasts than non-holders (Figures \ref{fig:error} and \ref{fig:error_backcasts}). 

This explanation implies that frequent price changes of commodities that directly affect households' finances likely raise their effort to acquire information on the macroeconomy given the sufficiently low cost of trading the commodity and acquiring the related information.  One prominent example is gasoline prices. Gasoline prices change daily, and the information is easily accessible, as local gas stations post the gasoline price at a location of high visibility. 
Gasoline is an essential part of household consumption directly affecting households' liquidity conditions regardless of stock ownership \citep{binder2022stuck}.\footnote{In MSC, about 97\% of stockholders and 90\% non-holders own or lease vehicles. Our results are robust when based on the subsample of respondents who own or lease vehicles. } 
If the frequency channel is in effect, we would expect that both stockholders and non-holders produce gasoline price forecasts of similar accuracy, unlike the accuracy of overall inflation forecasts.  

\begin{table}
    \centering
    \caption{Stock holding and Gas price forecast errors} \label{tab:forecast_control_gasoline}
    \small
    \centering
    \vspace{0.3cm}
	\noindent\resizebox{1.0\textwidth}{!}{
	\begin{tabularx}{\textwidth}{ll>{\centering}p{2.1cm}>{\centering}p{2.1cm}>{\centering}p{2.1cm}>{\centering}p{2.1cm}>{\centering}p{2.1cm}}
        \hline\hline
        \noalign{\vskip0.1cm}
         &  & (1) OLS & (2) IV & (3) FD & (3) OLS & (4) OLS \tabularnewline[0.1cm]
        \hline 
    	\noalign{\vskip0.1cm}
        Stock ($\beta$)  & &  0.181 &  0.181  &  -0.004 &  0.077  & 0.216  \tabularnewline
                    & &     (0.115)   & (0.347)    &   (0.322) &  (0.135)   & (0.135)   \tabularnewline[0.1cm]
        Attitude    & & -- & -- & -- &-1.218$^{***}$ & -- \tabularnewline
                    & &    & & & (0.074)  \tabularnewline[0.1cm]
        Sentiment    & & --  & --  & -- & --& -0.040$^{***}$  \tabularnewline
                    & &         & &  & &  (0.001)  \tabularnewline[0.1cm]
                    \hline 
        \noalign{\vskip0.1cm}
        Time FE          & & Y & Y & Y & Y   \tabularnewline[0.1cm]
        Demographics FE  & & Y & Y & Y & Y      \tabularnewline[0.1cm]
        Number of obs.  & & 64,492  &  26,980 & 27,111  & 67,460   & 67,460       \tabularnewline[0.1cm]
        $R^2$           & & 0.2866   &  -0.0001   &  0.4579 & 0.5475  & 0.5511   \tabularnewline[0.1cm]
        \hline\hline
        
    	\noalign{\vskip0.1cm}
    	\multicolumn{7}{p{\dimexpr\linewidth-2\tabcolsep-2\arrayrulewidth}}{\footnotesize \textit{Notes:} This table reports the regression results from Equations (\ref{eqn:error}) using gas price forecast errors as dependent variable. Columns (1) - (3) report estimation results using OLS, IV, and first-difference. Columns (3) and (4) report estimation results with attitude and sentiment as additional controls. ``Stock'' indicates dummies for stock market participation. We control for time-fixed effects and survey respondents' demographic fixed effects, including gender, education, birth cohort, marriage status, region, homeownership, and income quintiles. Robust standard errors are reported in the parenthesis.  $^{***}$, $^{**}$, $^{*}$ denotes statistical significance at 1\%, 5\%, and 10\% levels respectively.}    \\ 
    	\multicolumn{7}{p{\dimexpr\linewidth-2\tabcolsep-2\arrayrulewidth}}{\footnotesize \textit{Sources:} Authors' calculation.} 
	\end{tabularx}
    }
\end{table}

The empirical evidence indicates that this is indeed the case (Table \ref{tab:forecast_control_gasoline}). 
Stock-market participation does not create statistically significant differences in the accuracy of gasoline-price forecasts between stockholders and non-holders, quite distinguished from the accuracy of overall inflation forecasts.\footnote{Consistently, both groups' gas price forecast errors decrease to a rise in uncertainty, with no statistical difference between the two groups. Moreover, the share of gas price expenditure among non-stockholders might be larger than those of stockholders. Therefore, they have a higher incentive to pay attention to gas prices. This is in line with the positive coefficients of $\beta$'s in Table \ref{tab:forecast_control_gasoline}. However, such differences are not statistically significant.}  
Note that gasoline prices are largely determined by crude oil prices that change daily, while prices of other goods and services in the CPI basket do not change as frequently.  
The empirical evidence confirms the frequency channel, that is, households pay attention to frequent price changes that are directly relevant to household finance.

\subsection{The risk hedging channel} \label{ssec:uncertain} 

In general, stocks are risky assets, riskier than bonds and homes. 
According to \cite{huang2007rational}, economic agents with low-risk aversion invest in stocks, while those with high-risk aversion invest in bonds. 
Heterogeneity in risk aversion leads to different information choices: Stockholders choose frequent information with low accuracy, while bondholders choose less frequent information with high accuracy.

Distinguished from \cite{huang2007rational}, our empirical evidence points to the risk-hedging aspect of information acquisition which creates the endogenous feedback from stock holding. 
To illustrate, households acquire information on macroeconomic conditions for the timely detection of optimal selling (shorting) time to hedge against the high risk of stock holding. 
In this context, the cost associated with the information acquisition is analogous (or can be interpreted as) the flow cost of holding a put option or shorting a future of the stock, in that a large financial loss can be prevented by trading the stock at the right time. 
In other words, households internalize a hedging instrument by paying attention to the relevant macroeconomic news.

We empirically examine the risk-hedging channel. Since households want to reduce uncertainty or risk by raising attention, the risk-hedging motive increases with uncertainty.\footnote{Similarly, option prices rise with uncertainty according to the well-known Black-Scholes model of option pricing.} 
If the risk-hedging channel is in effect, we would see an increase in attention with rising uncertainty.  
To empirically verify this, consider the following model: 
\begin{equation}
    Y_{i,t} = \alpha + \beta_0 ~\text{\emph{Stock}}_{i,t} + \beta_1 ~\text{\emph{Stock}}_{i,t} \times \text{\emph{Uncertainty}}_t + \beta_2 ~\text{\emph{non-Stock}}_{i,t} \times \text{\emph{Uncertainty}}_t + \delta X_{i,t} + \epsilon_{i,t}, 
    \label{eqn:uncertain}
\end{equation}
where the dependent variable $Y_{i,t}$ is respondent $i$'s attention, or 1-year ahead inflation forecast error at time $t$. $\text{\emph{Stock}}_{i,t}$ ($\text{\emph{non-Stock}}_{i,t}$) is a dummy variable that equals $1$ ($0$) if respondent $i$ participates in stock market and $0$ ($1$) if not. 
For $\text{\emph{Uncertainty}}_t$, we consider two types of uncertainty measures that are most relevant to stock market performance: (1) market-based monetary policy uncertainty \citep{bauer2022market}, and (2) financial uncertainty \citep{ludvigson2021uncertainty}.\footnote{We aggregate the daily market-based monetary policy uncertainty measure to monthly by taking the maximum value over each month. We consider the 1-month ahead macro and financial uncertainty measures in \cite{ludvigson2021uncertainty} respectively. The results are robust to uncertainty measures over other horizons.}
The notation $X_{i,t}$ includes demographic fixed effects, including gender, education, birth cohort, marriage status, region, homeownership, and income quintiles. Results are reported in Table \ref{tab:uncertain}. As shown in Columns (1) and (2), when aggregate uncertainty rises, stockholders become more attentive than non-holders, as $\beta_2$ is larger than $\beta_1$ and the difference is statistically significant shown by the F-test statistics. 

\begin{table}
    \centering
    \caption{Sensitivity of attention to changes in economic uncertainty} \label{tab:uncertain}
    \small
    \centering
    \vspace{0.3cm}
	\noindent\resizebox{0.95\textwidth}{!}{
	\begin{tabularx}{\textwidth}{ll>{\centering}p{3.3cm}>{\centering}p{3.3cm}>{\centering}p{3.5cm}}
        \hline\hline
        \noalign{\vskip0.1cm}
        Uncertainty measure &  & (1) Monetary policy & (2) Financial & (3) Unemployment\tabularnewline[0.1cm]
        \hline 
    	\noalign{\vskip0.1cm}
        $\text{\emph{Stock}}$ ($\beta_0$) & &  7.35$^{***}$ & 2.87$^{**}$ & 7.407$^{***}$ \tabularnewline
                & & (0.83) & (1.44) & (0.56) \tabularnewline
        $\text{\emph{Stock}} \times \text{\emph{Uncertainty}}_t$ ($\beta_1$)  
                 & &  2.11$^{***}$ &  2.65$^{***}$  & 0.015$^{***}$\tabularnewline
                 & &     (0.17)    &    (0.14)      & (0.004)    \tabularnewline[0.1cm]
        $\text{\emph{non-Stock}} \times \text{\emph{Uncertainty}}_t$ ($\beta_2$)     
                 & &  1.25$^{***}$ &   1.45$^{***}$     & -0.003  \tabularnewline
                 & &     (0.27)    &   (0.21)      & (0.004) \tabularnewline[0.1cm]
                    \hline 
        \noalign{\vskip0.1cm}
        F-test: $\beta_1 = \beta_2$ & & 7.61$^{***}$ & 23.01$^{***}$  & 10.87$^{***}$\tabularnewline[0.1cm]
        Number of obs.              & & 120,147       &      120,147       & 47,735  \tabularnewline[0.1cm]
        Adj. $R^2$                       & & 0.0777        &    0.0794      & 0.1018    \tabularnewline[0.1cm]        
        \hline\hline
        
        \noalign{\vskip0.1cm}
        \multicolumn{5}{p{\dimexpr\linewidth-2\tabcolsep-2\arrayrulewidth}}{\footnotesize \textit{Notes:} This table reports the regression results from Equations (\ref{eqn:uncertain}). The dependent variable is $\textit{Attentiveness} \times 100$. ``Stock'' indicates dummies for stock market participation. ``Uncertainty'' denotes market-based monetary policy uncertainty (Column (1)), financial uncertainty (Column (2)), and individual probability of unemployment (Column (3)). The coefficients reported are based on a \emph{1-std.dev.} increase in the index for Columns (1) and (2). We control for survey respondents' demographic fixed effects, including gender, education, birth cohort, marriage status, region, homeownership, and income quintiles. In Column (3), we further control time-fixed effects. Robust standard errors are reported in the parenthesis.  $^{***}$, $^{**}$, $^{*}$ denotes statistical significance at 1\%, 5\%, and 10\% levels respectively.}   \\ 
        \multicolumn{5}{p{\dimexpr\linewidth-2\tabcolsep-2\arrayrulewidth}}{\footnotesize \textit{Sources:} Authors' calculation.} 
	\end{tabularx}
    }
\end{table}

In addition to the aggregate uncertainty indicator, we further consider the individual-level uncertainties. 
Specifically, the MSC has questions asking about an individual's unemployment probabilities over the next 5 years.\footnote{The question in MSC is phrased as ``During the next 5 years, what do you think the chances are that you (or your husband/wife) will lose a job you wanted to keep?''}
Households' answers to these questions reflect the subjective uncertainty about their labor market outcomes. When facing increasing chances of unemployment, stockholders may pay more attention to macroeconomic news to help hedge against such risks. 
Column (3) of Table \ref{tab:uncertain} strongly supports this hypothesis. When households' unemployment risk increases, stockholders hedge it by increasing their attentiveness to macroeconomic conditions. However, no significant changes are observed for non-holders in times of higher unemployment risk.\footnote{We find no significant differences between the subjective unemployment probabilities of stockholders and non-holders. Moreover, we have included time-fixed effects in the estimation with individual-level uncertainty to account for potential time-varying confounding factors.} 
To summarize, the risk-hedging channel is robust, regardless of aggregate or individual uncertainty measures.

\section{Policy Implications and Discussion} \label{sec:implications} 

This section explores the implications of our empirical findings on monetary policy and provides further corroborating evidence. Section \ref{sec:MP} discusses stock-driven attention can create a mechanism through which monetary policy has distributional consequences, favoring stockholding households over non-holders. Section \ref{sec:macronews} further supports this by highlighting that stockholders are better informed about overall macroeconomic news. 

\subsection{Implications to monetary policy} \label{sec:MP}

Through the lens of rational inattention theory, household attention is costly and endogenously determined in a way that maximizes household utility. Consequently, optimal attention allocation is potentially affected by factors such as their ex-ante risk aversion, stock market participation, homeownership, and the cost and benefit of information acquisition and trading assets in their portfolio. 
In particular, the three channels discussed in Section \ref{sec:mechanism} facilitate the feedback of stockholding to households' attention. Stock-market participation further raises the benefit of paying attention to macroeconomic conditions, which eventually enhances the precision of the acquired information. 
Consequently, stockholders optimally acquire more precise information despite the cost.

The attention heterogeneity driven by households' stock holding can create unintended consequences of monetary policy. First, increased stock-market participation and the resulting increase in households' attention to macroeconomic news may enhance the effectiveness of monetary policy.  During an economic recession when the aggregate economic uncertainty rises, households' attentiveness to economic news also increases.  This creates an environment where monetary policy, such as forward guidance, is effective, or perhaps more effective than in times of low uncertainty.  

Second, monetary policy may raise wealth inequality between stockholders and non-holders. In response to monetary policy shocks, stockholders, knowing the shock's effects on their financial situations, will re-optimize their consumption and investment more timely while non-holders are less likely to do so. 
The resulting welfare costs are smaller for stockholders, potentially exacerbating disparities between stock-market participants and nonparticipants.

We provide empirical evidence in support of this by showing the dynamic responses of the two groups' economic outlook to monetary policy shocks. 
We consider the effects of the orthogonalized monetary policy shock ($mps^{\perp}_t$) by \cite{bauer2023reassessment} versus the ``Fed response to news'' component, which is the difference between the unadjusted monetary policy surprise ($mps_t$) and orthogonalized monetary policy surprise.\footnote{Since the ``Fed response to news'' component summarizes the underestimation of how strongly the Fed would respond to economic news \citep{bauer2023alternative}, this component essentially provides a systematic summary of positive economic news, which we refer to as the ``demand news'' hereafter. An increase in this component indicates positive demand news.} Consider the following panel local projection: 
\begin{equation}
    Y_{i,t} = \alpha + \sum_{h=0}^{H}\beta_{1,h} ~\text{\emph{Stock}}_{i,t} \times \text{\emph{Shock}}_{t-h} + \sum_{h=0}^{H}\beta_{2,h} ~\text{\emph{non-Stock}}_{i,t} \times \text{\emph{Shock}}_{t-h} + \delta X_{i,t} + \epsilon_{i,t}, 
    \label{eqn:dynamic}
\end{equation}
where the dependent variable $Y_{i,t}$ denotes respondent $i$'s inflation expectation, business conditions outlook, or sentiment at time $t$. 
The notation $X_{i,t}$ captures demographic fixed effects such as gender, education, birth cohort, marriage status, region, homeownership, and income quintiles. 
Coefficients $\beta_{1,h}$ and $\beta_{2,h}$ represent the effect of the shock on stockholders' and non-stockholders' expectations over $h$ months. 

\begin{figure}
    \centering
    \caption{Impulse responses to monetary policy news}
    \includegraphics[width = 0.45\textwidth]{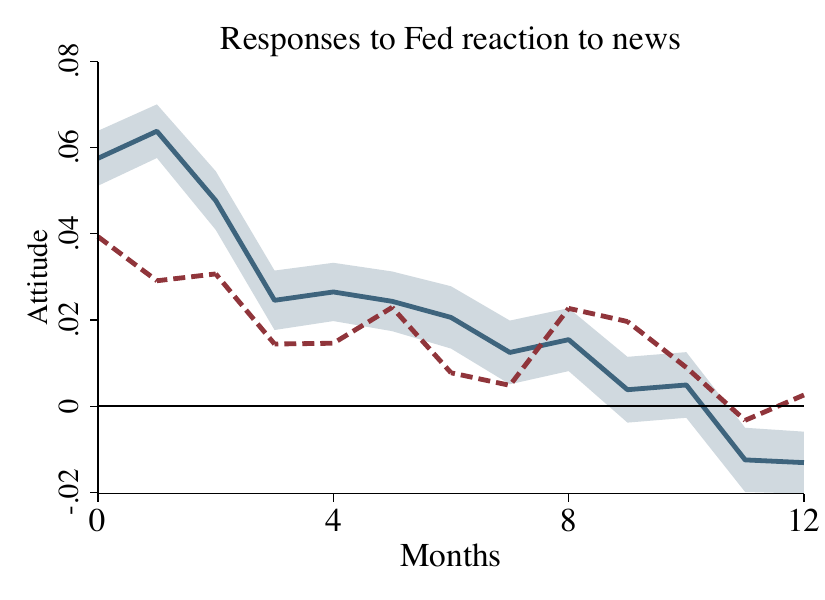}
    \includegraphics[width = 0.45\textwidth]{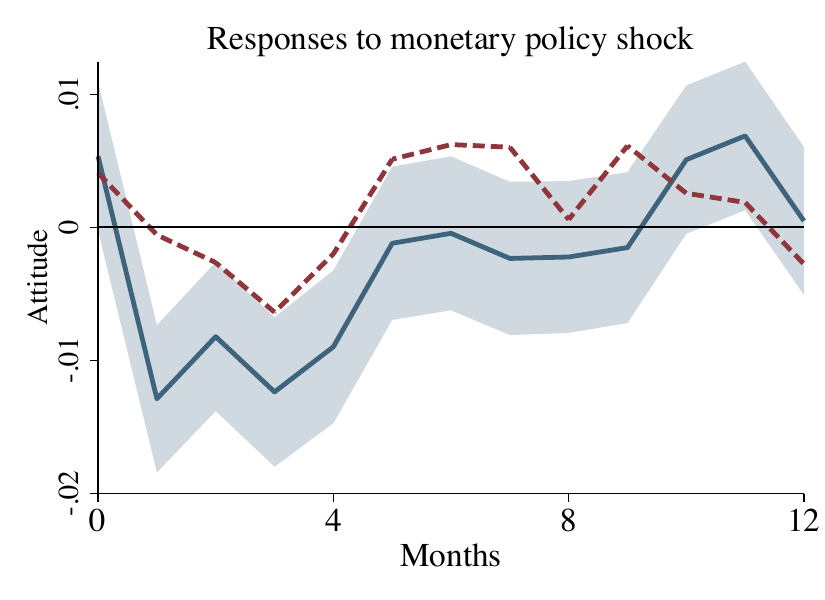}
    \includegraphics[width = 0.45\textwidth]{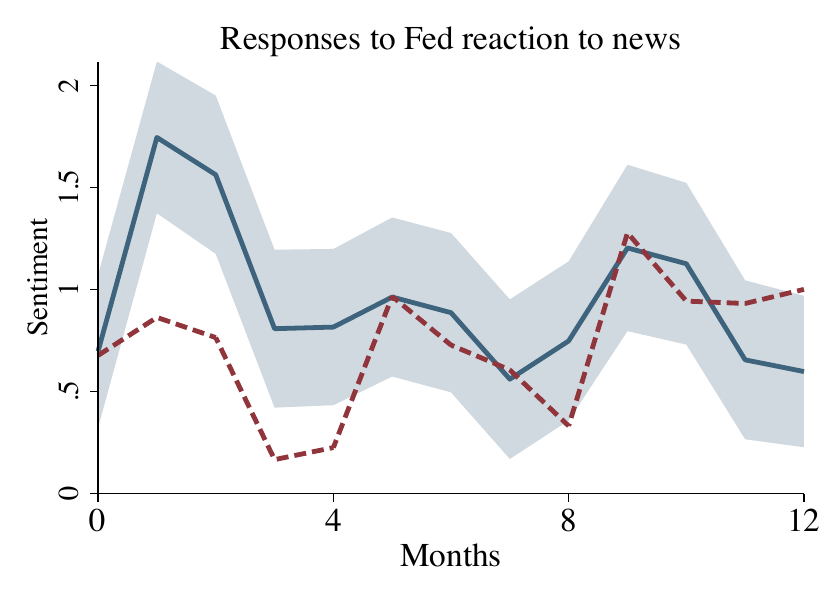}
    \includegraphics[width = 0.45\textwidth]{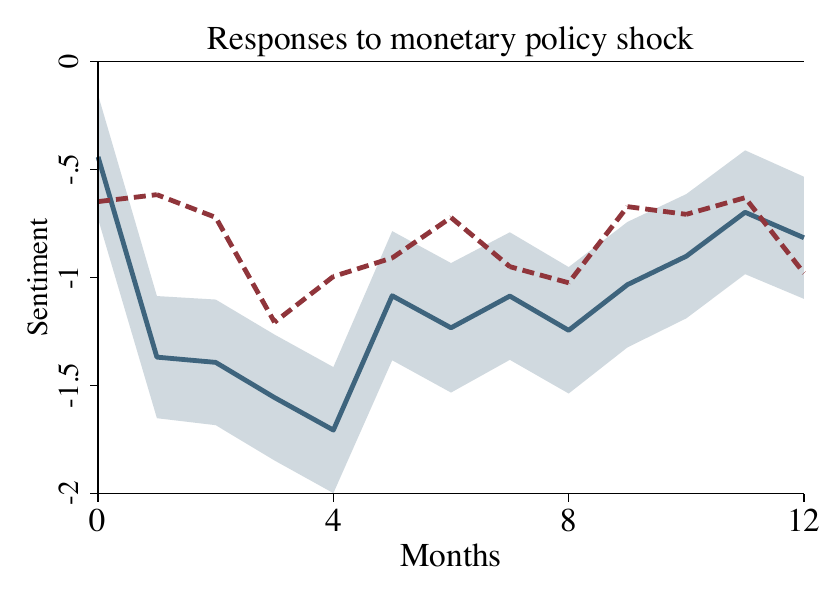}
    \includegraphics[width = 0.45\textwidth]{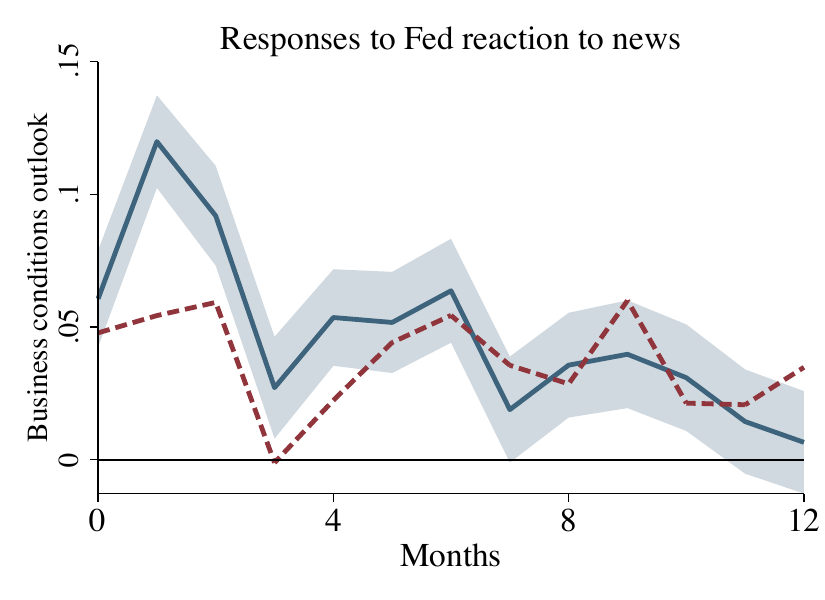}
    \includegraphics[width = 0.45\textwidth]{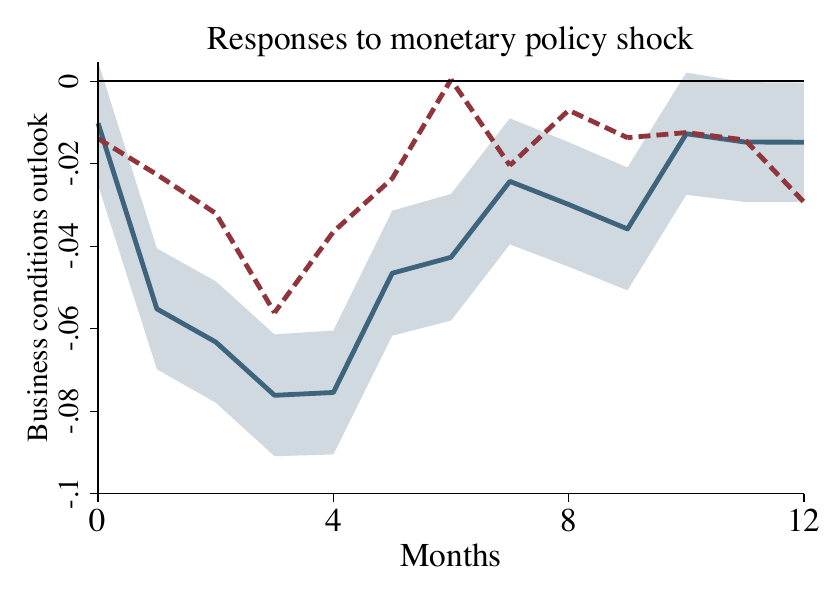}
    \label{fig:irf}

    \noindent\resizebox{1.0\textwidth}{!}{
		\begin{tabularx}{\textwidth}{m{3.8cm}}
			\noalign{\vskip0.3cm}
			\multicolumn{1}{p{\dimexpr\linewidth-2\tabcolsep-2\arrayrulewidth}}{
				\footnotesize \textit{Notes:} This figure plots the impulse response of stockholders' (solid blue line) and non-holders' (dashed red line) expectations to monetary policy news. The left and right columns are based on the Fed's response to the news and monetary policy surprises respectively. Dependent variables from top to bottom are (1) attitude, (2) sentiment, and (3) business condition outlooks. Shaded areas are 95\% CIs.} \\
			\multicolumn{1}{p{\dimexpr\linewidth-2\tabcolsep-2\arrayrulewidth}}{	
				\footnotesize \textit{Source:} Authors' calculation. 
            }
            \end{tabularx}
    }
\end{figure}

The left column in Figure \ref{fig:irf} displays households' responses to the demand news. On average, stockholders show a more optimistic attitude (top panel), sentiment (second panel from the top), and business conditions outlook (third panel from the top) than non-holders do. The right column displays households' reactions to contractionary monetary policy shock from \cite{bauer2023reassessment}. Stockholders lower their overall economic outlook and assessment more than nonholders do, and the difference is statistically different. Our results support our claim that monetary policy may have stronger effects on stockholders' beliefs due to their higher attentiveness to macroeconomic conditions.

\subsection{Further discussions} \label{sec:macronews}

We provide further empirical evidence in support of the distributional effects by showing the responses of the two groups' economic outlook to other structural macroeconomic shocks. To this end, we examine whether stockholders understand or perceive the nature of various economic shocks in a way consistent with the prediction of commonly believed macroeconomic theories. 

We examine the responsiveness of households' economic outlook to both demand- and supply-side macroeconomic news shocks. We focus on news shocks as expectations are forward-looking variables. 
Specifically, we employ the Total Factor Productivity (TFP) news shock by \cite{barsky2011news} as our example of supply news shock and the ``Fed response to news'' during the month $t$ \citep{bauer2023alternative} as our example of demand news shock.

Consider the following regression model: 
\begin{equation}
    Y_{i,t} = \alpha + \beta_1 ~\text{\emph{Stock}}_{i,t} \times \text{\emph{Shock}}_t + \beta_2 ~\text{\emph{non-Stock}}_{i,t} \times \text{\emph{Shock}}_t + \delta X_{i,t} + \epsilon_{i,t}, 
    \label{eqn:shock}
\end{equation}
where the dependent variable $Y_{i,t}$ denotes respondent $i$'s inflation expectation, business conditions outlook, or sentiment at time $t$. The business conditions outlook is re-coded as a categorical variable ranging from 1 - 5, with 1 denoting ``Bad Times'' and 5 denoting ``Good Times''. Therefore, an increase in the value captures improvements in the business conditions outlooks. The notation $X_{i,t}$ are demographic fixed effects, including gender, education, birth cohort, marriage status, region, homeownership, and income quintiles; $\text{\emph{Shock}}_t$ is one of the macroeconomic shocks discussed.

The empirical results are reported in Table \ref{tab:shock}. 
Panel A reports the responsiveness to demand news -- ``Fed response to news''. 
Given positive demand news, households' inflation expectations rise, and both business conditions outlook and consumer sentiment improve with statistical significance. 
Meanwhile, non-holders' inflation expectations show limited responsiveness with the coefficient ($\beta_2$) being not statistically significant. Their outlook on business conditions and economic sentiment also show smaller reactions than those of stockholders. 
Again, stockholders are likely better informed of FOMC's take on recent data releases and adjust their economic outlook accordingly. 

Panel B reports the responsiveness to a positive TFP news shock. 
Stockholders' inflation expectations decline, while their outlook on business conditions outlook and economic sentiment improve. 
These reactions---all statistically significant---are consistent with the positive supply-shock nature of TFP news shock. 
Meanwhile, non-holders' inflation expectations do not decline meaningfully with a statistically insignificant coefficient ($\beta_2$). 
This evidence suggests that stockholders are better informed of the supply-shock nature of TFP news shocks and adjust their macroeconomic expectations accordingly. 

\begin{table}[ht!]
    \centering
    \caption{Sensitivity of households' expectations to macroeconomic shocks} \label{tab:shock}
    \small
    \centering
    \vspace{0.3cm}
	\noindent\resizebox{1.0\textwidth}{!}{
	\begin{tabularx}{\textwidth}{ll>{\centering}p{3.5cm}>{\centering}p{3.5cm}>{\centering}p{3.5cm}}
        \hline\hline
        \noalign{\vskip0.1cm}
        Macroeconomic shocks &  & (1) Inflation & (2) Business conditions & (3) Sentiment \tabularnewline[0.1cm]
        \hline 
        \noalign{\vskip0.1cm}
        \multicolumn{5}{l}{\textbf{\textit{Panel A. Fed reaction to news}}} \tabularnewline
        \noalign{\vskip0.1cm}
        $\text{\emph{Stock}} \times \text{\emph{Shock}}_t$ ($\beta_1$)  
                 & &   0.134$^{*}$ &  0.188$^{***}$  &   3.127$^{***}$  \tabularnewline
                 & &     (0.020)    &    (0.008)      &    (0.163)          \tabularnewline[0.1cm]
        $\text{\emph{non-Stock}} \times \text{\emph{Shock}}_t$ ($\beta_2$)     
                 & &  0.071        &  0.144$^{***}$ &   2.534$^{***}$       \tabularnewline
                 & &     (0.035)    &   (0.011)       &    (0.216)   \tabularnewline[0.1cm]
        \noalign{\vskip0.1cm}
        \hline 
        \noalign{\vskip0.1cm}
        F-test: $\beta_1 = \beta_2$ & & 2.79$^{*}$   &     10.14$^{***}$          &  4.83$^{**}$ \tabularnewline[0.1cm]
        Number of obs.              & & 108,755       &     110.523      &  120,147      \tabularnewline[0.1cm]
        $R^2$                       & & 0.0285        &    0.0294       &  0.0768     \tabularnewline[0.1cm]
        \noalign{\vskip0.1cm}
        \hline 
        
        \noalign{\vskip0.1cm}
        \multicolumn{5}{l}{\textbf{\textit{Panel B. TFP news shocks}}} \tabularnewline
        \noalign{\vskip0.1cm}
        $\text{\emph{Stock}} \times \text{\emph{Shock}}_t$ ($\beta_1$)  
                 & &   -0.044$^{*}$ &  0.162$^{***}$  &   3.167$^{***}$  \tabularnewline
                 & &     (0.024)    &    (0.013)      &    (0.242)          \tabularnewline[0.1cm]
        $\text{\emph{non-Stock}} \times \text{\emph{Shock}}_t$ ($\beta_2$)     
                 & & -0.003         &  0.187$^{***}$ &   2.559$^{***}$       \tabularnewline
                 & &     (0.035)    &   (0.016)       &    (0.275)   \tabularnewline[0.1cm]
        \noalign{\vskip0.1cm}
        \hline 
        \noalign{\vskip0.1cm}
        F-test: $\beta_1 = \beta_2$ & & 0.93   &     1.6          &  2.78$^{*}$ \tabularnewline[0.1cm]
        Number of obs.              & & 45,934       &     47,868      &  51,330      \tabularnewline[0.1cm]
        $R^2$                       & & 0.0259        &    0.0316       &  0.0878     \tabularnewline[0.1cm]
        \noalign{\vskip0.1cm}

        \hline\hline
        
        \noalign{\vskip0.1cm}
        \multicolumn{5}{p{\dimexpr\linewidth-2\tabcolsep-2\arrayrulewidth}}{\footnotesize \textit{Notes:} This table reports the regression results from Equations (\ref{eqn:shock}). The dependent variables are 1-year ahead inflation expectations (Column (1)), business conditions outlook (Column (2)), and index of consumer sentiment (Column (3)). ``Stock'' indicates dummies for stock market participation. ``Shock'' denotes Fed information shock (Panel A), TFP news shock (Panel B), and Fed reaction to news (Panel C). The coefficients reported are based on a \emph{1-std.dev.} increase in each type of shock. We control for survey respondents' demographic fixed effects, including gender, education, birth cohort, marriage status, region, homeownership, and income quintiles. Robust standard errors are reported in the parenthesis.  $^{***}$, $^{**}$, $^{*}$ denotes statistical significance at 1\%, 5\%, and 10\% levels respectively.}   \\ 
        {\footnotesize \textit{Sources:} Authors' calculation.} 
	\end{tabularx}
    }
\end{table}


Next, we consider the dynamic responses of inflation expectations to news shocks on demand and supply.
Households adjust their inflation expectations quite differently to supply and demand shocks, but the responsiveness is stronger among stockholders than among non-holders. As shown in the left column in Figure \ref{fig:irf_inflation}, both stockholders and non-holders lower their inflation expectations to a positive TFP shock, consistent with the supply shock nature of TFP shock. Note that stockholders lower inflation expectations more than non-holders do, suggesting that stockholders have a better understanding of the shock and adjust the inflation outlook accordingly. The two groups raised their inflation outlook after observing the Fed's interpretation of recent economic news, consistent with the demand shock nature of the shock. Stockholders raise their inflation outlook more than non-holders do, but the difference is not statistically significant. 

All told, in the presence of structural shocks, stockholders are more likely to understand the nature of shocks and adjust their macroeconomic expectations in a way that is consistent with predictions of standard macroeconomic models. However, the responsiveness of non-holders is limited.

\section{Conclusions} \label{sec:conclusion}

We conclude by discussing implications for future research. Our empirical findings suggest the importance of households' stock investment and financial conditions in their attention to the macroeconomy and expectation formation. This paper focuses only on stock-market participation due to data availability, but a comprehensive investigation using individual-level data on detailed asset and debt holdings would lead us to a more complete understanding of the link between household finance and expectation formation. Furthermore, our empirical findings shed light on the endogenous attention channel driven by households' portfolios, which has not been well studied in a structural model. This agenda can be pursued in future research.

%
%

\singlespacing
{{
\setlength{\bibsep}{.2cm}
\bibliographystyle{aer}
\bibliography{references}

@article{coibion2015information,
  title={Information rigidity and the expectations formation process: A simple framework and new facts},
  author={Coibion, Olivier and Gorodnichenko, Yuriy},
  journal={American Economic Review},
  volume={105},
  number={8},
  pages={2644--2678},
  year={2015},
  publisher={American Economic Association 2014 Broadway, Suite 305, Nashville, TN 37203}
}

@TechReport{BrachaTang2022,
  author={Anat Bracha and Jenny Tang},
  title={{Inflation Levels and (In)Attention}},
  year=2022,
  month=Jan,
  institution={Federal Reserve Bank of Boston},
  type={Working Papers},
  url={https://ideas.repec.org/p/fip/fedbwp/93857.html},
  number={22-4},
  doi={10.29412/res.wp.2022.04},
}

@article{upbias_jep,
author = {Weber, Michael and D'Acunto, Francesco and Gorodnichenko, Yuriy and Coibion, Olivier},
title = {The Subjective Inflation Expectations of Households and Firms: Measurement, Determinants, and Implications},
journal = {Journal of Economic Perspectives},
volume = { },
number = { },
pages = { },
doi = { },
year = {Forthcoming}
}

@TechReport{Axelrod2018,
  author={Sandor Axelrod and David E. Lebow and Ekaterina V. Peneva},
  title={{Perceptions and Expectations of Inflation by U.S. Households}},
  year=2018,
  month=Oct,
  institution={Board of Governors of the Federal Reserve System (U.S.)},
  type={Finance and Economics Discussion Series},
  url={https://ideas.repec.org/p/fip/fedgfe/2018-73.html},
  number={2018-073},
  doi={10.17016/FEDS.2018.073},
}

@article{bauer2023alternative,
  title={An alternative explanation for the “fed information effect”},
  author={Bauer, Michael D and Swanson, Eric T},
  journal={American Economic Review},
  volume={113},
  number={3},
  pages={664--700},
  year={2023},
  publisher={American Economic Association 2014 Broadway, Suite 305, Nashville, TN 37203}
}

@article{CG2012JPE,
 ISSN = {00223808, 1537534X},
 URL = {http://www.jstor.org/stable/10.1086/665662},
 author = {Olivier Coibion and Yuriy Gorodnichenko},
 journal = {Journal of Political Economy},
 number = {1},
 pages = {116--159},
 publisher = {The University of Chicago Press},
 title = {What Can Survey Forecasts Tell Us about Information Rigidities?},
 urldate = {2023-11-22},
 volume = {120},
 year = {2012}
}

@article{barsky2011news,
  title={News shocks and business cycles},
  author={Barsky, Robert B and Sims, Eric R},
  journal={Journal of Monetary Economics},
  volume={58},
  number={3},
  pages={273--289},
  year={2011},
  publisher={Elsevier}
}

@article{ludvigson2021uncertainty,
  title={Uncertainty and business cycles: exogenous impulse or endogenous response?},
  author={Ludvigson, Sydney C and Ma, Sai and Ng, Serena},
  journal={American Economic Journal: Macroeconomics},
  volume={13},
  number={4},
  pages={369--410},
  year={2021}
}

@article{bauer2023reassessment,
  title={A reassessment of monetary policy surprises and high-frequency identification},
  author={Bauer, Michael D and Swanson, Eric T},
  journal={NBER Macroeconomics Annual},
  volume={37},
  number={1},
  pages={87--155},
  year={2023},
  publisher={The University of Chicago Press Chicago, IL}
}

@article{bauer2022market,
  title={Market-based monetary policy uncertainty},
  author={Bauer, Michael D and Lakdawala, Aeimit and Mueller, Philippe},
  journal={The Economic Journal},
  volume={132},
  number={644},
  pages={1290--1308},
  year={2022},
  publisher={Oxford University Press}
}

@article{luo2010rational,
  title={Rational inattention, long-run consumption risk, and portfolio choice},
  author={Luo, Yulei},
  journal={Review of Economic dynamics},
  volume={13},
  number={4},
  pages={843--860},
  year={2010},
  publisher={Elsevier}
}

@article{das2019,
    author = {Das, Sreyoshi and Kuhnen, Camelia M and Nagel, Stefan},
    title = "{Socioeconomic Status and Macroeconomic Expectations}",
    journal = {The Review of Financial Studies},
    volume = {33},
    number = {1},
    pages = {395-432},
    year = {2019},
    month = {03},
    issn = {0893-9454},
    doi = {10.1093/rfs/hhz041},
    url = {https://doi.org/10.1093/rfs/hhz041},
    eprint = {https://academic.oup.com/rfs/article-pdf/33/1/395/32742335/hhz041.pdf},
}

@article{huang2007rational,
  title={Rational inattention and portfolio selection},
  author={Huang, Lixin and Liu, Hong},
  journal={The Journal of Finance},
  volume={62},
  number={4},
  pages={1999--2040},
  year={2007},
  publisher={Wiley Online Library}
}

@TechReport{Mitman2021,
  author={Mitman, Kurt and Broer, Tobias and Kohlhas, Alexandre and Schlafmann, Kathrin},
  title={{Information and Wealth Heterogeneity in the Macroeconomy}},
  year=2021,
  month=Mar,
  institution={C.E.P.R. Discussion Papers},
  type={CEPR Discussion Papers},
  url={https://ideas.repec.org/p/cpr/ceprdp/15934.html},
  number={15934},
  doi={},
}

@TechReport{LiSinha2023,
  author={Geng Li and Nitish Ranjan Sinha},
  title={{Are Real Assets Owners Less Averse to Inflation? Evidence from Consumer Sentiments and Inflation Expectations}},
  year=2023,
  month=August,
  institution={Board of Governors of the Federal Reserve System (U.S.)},
  type={Finance and Economics Discussion Series},
  url={https://www.federalreserve.gov/econres/feds/are-real-assets-owners-less-averse-to-inflation-evidence-from-consumer-sentiments-and-inflation-expectations.htm},
  number={2023-058},
  doi={10.17016/FEDS.2023.058},
}

@article{AXY,
  title={Effects of monetary policy on household expectations: The role of homeownership},
  author={Ahn, Hie Joo and Xie, Shihan and Yang, Choongryul},
  journal={Journal of Monetary Economics},
  pages={103599},
  year={2024},
  publisher={Elsevier}
}

@article{Xie2023,
    author = {Xie, Shihan},
    title = "{An Estimated Model of Household Inflation Expectations: Information Frictions and Implications}",
    journal = {The Review of Economics and Statistics},
    year = {forthcoming},
    issn = {0034-6535},
    doi = {10.1162/rest_a_01318},
    url = {https://doi.org/10.1162/rest\_a\_01318},
    eprint = {https://direct.mit.edu/rest/article-pdf/doi/10.1162/rest\_a\_01318/2075028/rest\_a\_01318.pdf},
}

@article{link2024attention,
  title={Attention to the Macroeconomy},
  author={Link, Sebastian and Peichl, Andreas and Roth, Christopher and Wohlfart, Johannes},
  journal={Available at SSRN 4697814},
  year={2024}
}

@article{binder2017measuring,
  title={Measuring uncertainty based on rounding: New method and application to inflation expectations},
  author={Binder, Carola C},
  journal={Journal of Monetary Economics},
  volume={90},
  pages={1--12},
  year={2017},
  publisher={Elsevier}
}

@article{kuhnen2017socioeconomic,
  title={Socioeconomic status and learning from financial information},
  author={Kuhnen, Camelia M and Miu, Andrei C},
  journal={Journal of Financial Economics},
  volume={124},
  number={2},
  pages={349--372},
  year={2017},
  publisher={Elsevier}
}

@article{kuhnen2015asymmetric,
  title={Asymmetric learning from financial information},
  author={Kuhnen, Camelia M},
  journal={The Journal of Finance},
  volume={70},
  number={5},
  pages={2029--2062},
  year={2015},
  publisher={Wiley Online Library}
}

@article{itzhak2018,
 title = "Extrapolative Uncertainty and Household Economic Behavior",
 author = "Ben-David, Itzhak and Fermand, Elyas and Kuhnen, Camelia M and Li, Geng",
 journal = "Management Science",
 year = "2024",
}

@article{drager2017imperfect,
  title={Imperfect information and consumer inflation expectations: Evidence from microdata},
  author={Dr{\"a}ger, Lena and Lamla, Michael J},
  journal={Oxford Bulletin of Economics and Statistics},
  volume={79},
  number={6},
  pages={933--968},
  year={2017},
  publisher={Wiley Online Library}
}

@article{andrade2019forward,
  title={Forward guidance and heterogeneous beliefs},
  author={Andrade, Philippe and Gaballo, Gaetano and Mengus, Eric and Mojon, Benoit},
  journal={American Economic Journal: Macroeconomics},
  volume={11},
  number={3},
  pages={1--29},
  year={2019},
  publisher={American Economic Association 2014 Broadway, Suite 305, Nashville, TN 37203-2425}
}

@article{kim2023learning,
  title={Learning-through-survey in inflation expectations},
  author={Kim, Gwangmin and Binder, Carola},
  journal={American Economic Journal: Macroeconomics},
  volume={15},
  number={2},
  pages={254--278},
  year={2023},
  publisher={American Economic Association 2014 Broadway, Suite 305, Nashville, TN 37203-2425}
}

@article{cloyne2020monetary,
  title={Monetary policy when households have debt: new evidence on the transmission mechanism},
  author={Cloyne, James and Ferreira, Clodomiro and Surico, Paolo},
  journal={The Review of Economic Studies},
  volume={87},
  number={1},
  pages={102--129},
  year={2020},
  publisher={Oxford University Press}
}

@article{luetticke2021transmission,
  title={Transmission of monetary policy with heterogeneity in household portfolios},
  author={Luetticke, Ralph},
  journal={American Economic Journal: Macroeconomics},
  volume={13},
  number={2},
  pages={1--25},
  year={2021},
  publisher={American Economic Association 2014 Broadway, Suite 305, Nashville, TN 37203-2425}
}

@article{luo2017robustly,
  title={Robustly strategic consumption--portfolio rules with informational frictions},
  author={Luo, Yulei},
  journal={Management Science},
  volume={63},
  number={12},
  pages={4158--4174},
  year={2017},
  publisher={INFORMS}
}

@article{binder2022stuck,
  title={Stuck in the seventies: gas prices and consumer sentiment},
  author={Binder, Carola and Makridis, Christos},
  journal={Review of Economics and Statistics},
  volume={104},
  number={2},
  pages={293--305},
  year={2022},
  publisher={MIT Press One Rogers Street, Cambridge, MA 02142-1209, USA journals-info~…}
}
}}

%
%
\clearpage

\onehalfspacing

\renewcommand\appendix{\par
\newcounter{section1}
\setcounter{section}{0}
\setcounter{subsection}{0}
\setcounter{equation}{0}
\setcounter{table}{0}
\setcounter{figure}{0}
\setcounter{page}{1}
\gdef\thesection{Appendix \Alph{section}}
\gdef\thefigure{\Alph{section}.\arabic{figure}}
\gdef\theequation{\Alph{section}.\arabic{equation}}
\gdef\thetable{\Alph{section}.\arabic{table}}
}

\renewcommand{\figurename}{Appendix Figure}
\renewcommand{\tablename}{Appendix Table}

\renewcommand{\theHsection}{appendixsection.\Alph{section}}

\newpage 
\appendix

\begin{center}
    \textbf{\LARGE Appendix For Online Publication}
\end{center}

\section{Attentiveness and attitudes} 

\subsection{The attentiveness and attitude indices}\label{app:index}

\begin{figure}[ht!]
    \centering
    \includegraphics[width = 0.85\textwidth]{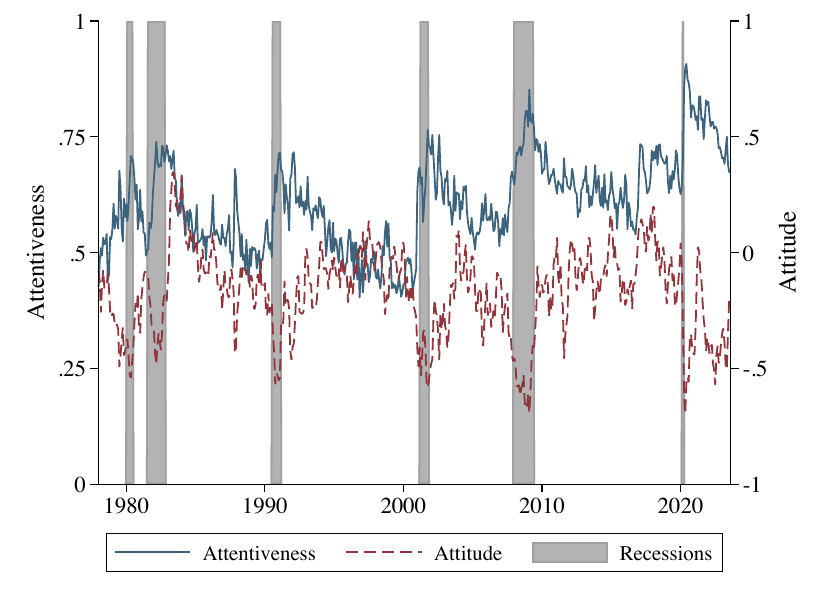}
    \caption{Time series plots of attentiveness and attitude} 
    \label{fig:attention}
	\noindent\resizebox{1.0\textwidth}{!}{
		\begin{tabularx}{\textwidth}{m{3.8cm}}
			\noalign{\vskip0.3cm}
			\multicolumn{1}{p{\dimexpr\linewidth-2\tabcolsep-2\arrayrulewidth}}{
				\footnotesize \textit{Notes:} This figure plots the authors' constructed measures of aggregate households' \emph{attentiveness} (solid blue) and \emph{attitude} (dotted red) towards economic conditions. Shaded areas are NBER recession dates. The sample period is 1978M1-2023M8.} \\
			\multicolumn{1}{p{\dimexpr\linewidth-2\tabcolsep-2\arrayrulewidth}}{	
				\footnotesize \textit{Source:} Authors' calculation. 
            }
		\end{tabularx}
	}
\end{figure} 

With the individual-level indicators, we further construct aggregate-level measures of households' attentiveness and attitude, which we call \emph{``the attentiveness index''} and \emph{``the attitude index''}, respectively. Figure \ref{fig:attention} depicts the indices of attentiveness and attitude. The two indices show quite different dynamics. The attentiveness index is countercyclical whereas the attitude index is procyclical. Households pay more attention to economic news when macroeconomic conditions deteriorate. The attitude index indicates that their memory of recent macroeconomic events is consistent with the realized cyclical effects of these events on the aggregate economy. The opposite cyclical properties of the two indices suggest that more negative recalls largely drive increased attentiveness during an economic downturn. Households may become more attentive to economic news as uncertainty increases during recessions.

\subsection{Attentiveness, attitude, and inflation expectations} \label{app:att}
We then document the relationship between individual-level attentiveness and attitude with inflation forecast. We regress households' expected inflation on attentiveness and attitude indicators while controlling for individual attributes and time-fixed effects. Column (1) in Table \ref{tab:attitude_fe} reports the estimation result. As documented in the previous literature (\citealp{upbias_jep}), there is an upward bias in households' inflation expectations, i.e., households' 1-year-ahead inflation expectation is higher than the realized CPI inflation. On average, attentive households report lower and more accurate inflation forecasts. 

\begin{table}
    \centering
    \caption{Attentiveness, Attitude, and Inflation Forecasts} \label{tab:attitude_fe}
    \small
    \centering
    \vspace{0.3cm}
	\noindent\resizebox{0.95\textwidth}{!}{
	\begin{tabularx}{\textwidth}{l>{\centering}p{3.5cm}>{\centering}p{3.5cm}>{\centering}p{3.5cm}}
        \hline\hline
        \noalign{\vskip0.1cm}
    Dependent variable   & (1) Level & (2) Uncertainty & (3) Inattention \tabularnewline
    \noalign{\vskip0.1cm}
    \hline 
    \noalign{\vskip0.1cm}
    Attentiveness          &       -0.094$^{***}$ &   -0.018$^{***}$ & -0.019$^{***}$ \tabularnewline
                             &     (0.025)        &   (0.002) &  (0.006) \tabularnewline
    Attitude         &        -0.561$^{***}$ &    -0.024$^{***}$ & 0.002 \tabularnewline
                     &     (0.016)       & (0.001) & (0.04) \tabularnewline
    \noalign{\vskip0.1cm}
    Demographic FE  & $\checkmark$   & $\checkmark$ & $\checkmark$ \tabularnewline
    Time   FE       & $\checkmark$   & $\checkmark$ & $\checkmark$ \tabularnewline
    \hline 
    \noalign{\vskip0.1cm}
    adj. $ R^2$         & 0.0698     & 0.1085  & 0.0632     \tabularnewline
    Observations        & 109,115     & 108,859   & 14,211   \tabularnewline
    \noalign{\vskip0.1cm}
    \hline \hline 
    	\noalign{\vskip0.1cm}
    	\multicolumn{4}{p{\dimexpr\linewidth-2\tabcolsep-2\arrayrulewidth}}{\footnotesize \textit{Notes:} The dependent variables in Columns (1) - (3) are inflation forecasts, uncertainty, and inattention respectively. Demographic fixed effects include stock market participation, homeownership, cohort, education, income quintile, gender, and region. Robust standard errors are reported. The sample period is 1978M1-2023M8. $^{***}$, $^{**}$, $^{*}$ denotes statistical significance at 1\%, 5\%, and 10\% levels respectively.}    \tabularnewline
    	{\footnotesize \textit{Sources:} Authors' calculation.} 
	\end{tabularx}
    }
\end{table}

Next, we investigate the properties of our measures of attentiveness and attitudes with other related individual-level indices developed using MSC. \cite{binder2017measuring} develops a method to measure individual-level uncertainty based on rounding. We replicate this uncertainty measure and show that more attentive individuals are also more certain about their inflation expectations in Column (2) of Table \ref{tab:attitude_fe}. 
\cite{BrachaTang2022} constructs a measure of inflation inattention relying on consumers’ current inflation estimates. Following their measure, individuals who believe inflation during the next 12 months is the same as the current inflation level while unable to provide a numeric estimate of current inflation are labeled as inattentive. Column (3) of Table \ref{tab:attitude_fe} shows that individuals who are in general attentive to macroeconomic conditions are less likely to be inattentive to current inflation. 

In sum, attentive individuals, everything else equal, report lower inflation forecasts. They are also less likely to be uncertain about their inflation forecasts and are more likely to be knowledgeable about the current level of inflation. These observations suggest that our new indicator of attentiveness effectively captures households' information acquisition efforts.

\section{Additional tables and figures}

\begin{figure}[ht!]
    \centering
    \includegraphics[width = 0.55\textwidth]{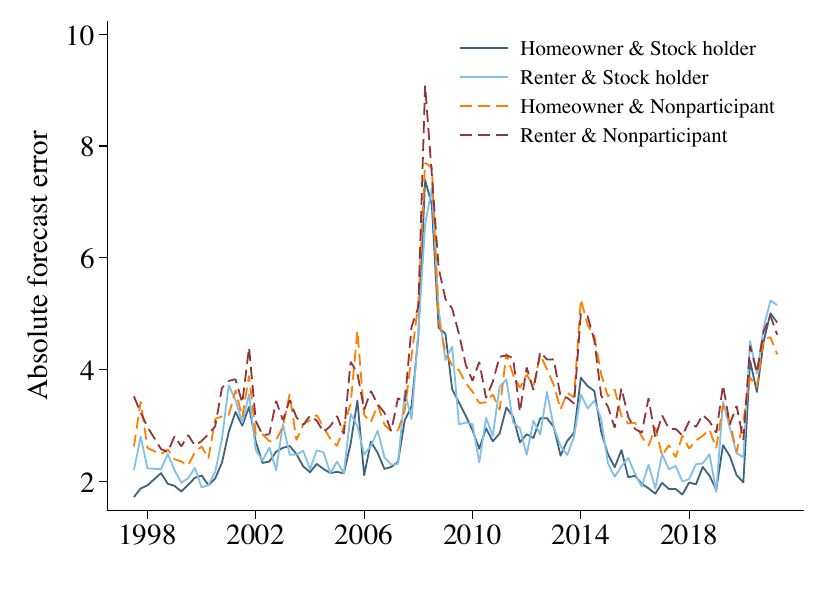}
    \\
    \includegraphics[width = 0.55\textwidth]{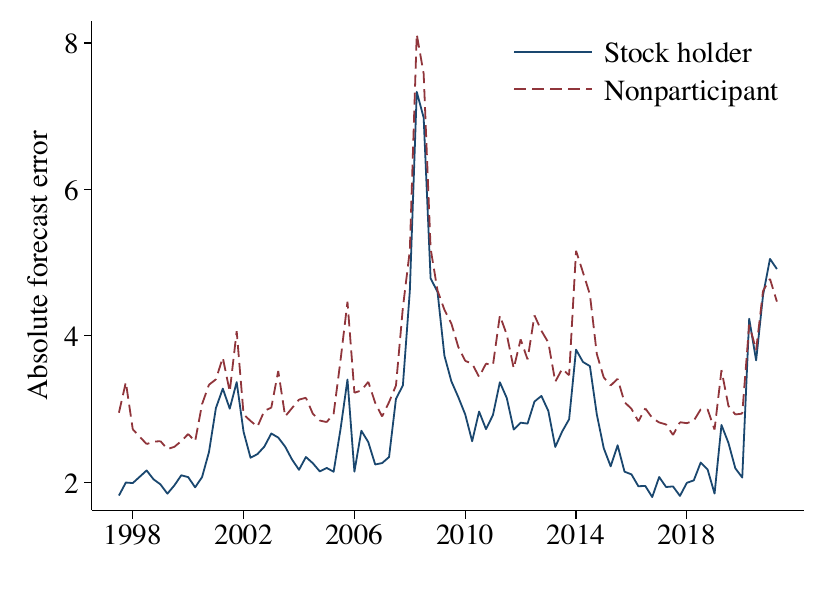}
    \\
    \includegraphics[width = 0.55\textwidth]{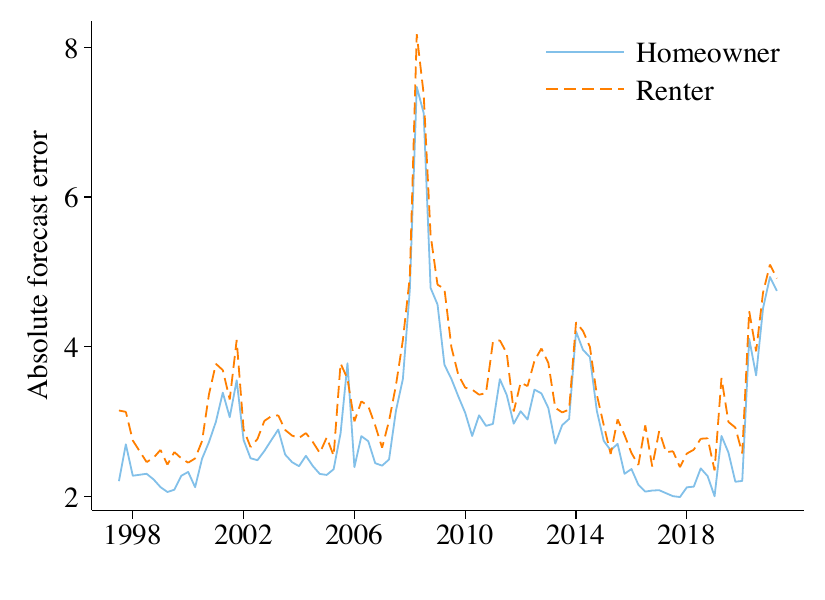}
    \caption{Forecast accuracy by asset-holding status} 
    \label{fig:error}
    
	\noindent\resizebox{1.0\textwidth}{!}{
		\begin{tabularx}{\textwidth}{m{3.8cm}}
			\noalign{\vskip0.3cm}
			\multicolumn{1}{p{\dimexpr\linewidth-2\tabcolsep-2\arrayrulewidth}}{
				\footnotesize \textit{Note:} The average absolute forecast errors are reported.} \\
			\multicolumn{1}{p{\dimexpr\linewidth-2\tabcolsep-2\arrayrulewidth}}{	
				\footnotesize \textit{Source:} Authors' calculation. 
            }
		\end{tabularx}
	}
\end{figure}

\begin{figure}[ht!]
    \centering
    \includegraphics[width = 0.55\textwidth, height = 0.45\textwidth]{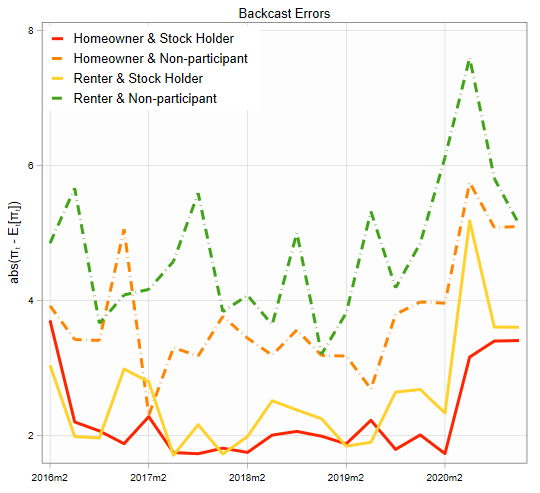} \\ 
    \includegraphics[width = 0.55\textwidth, height = 0.7\textwidth]{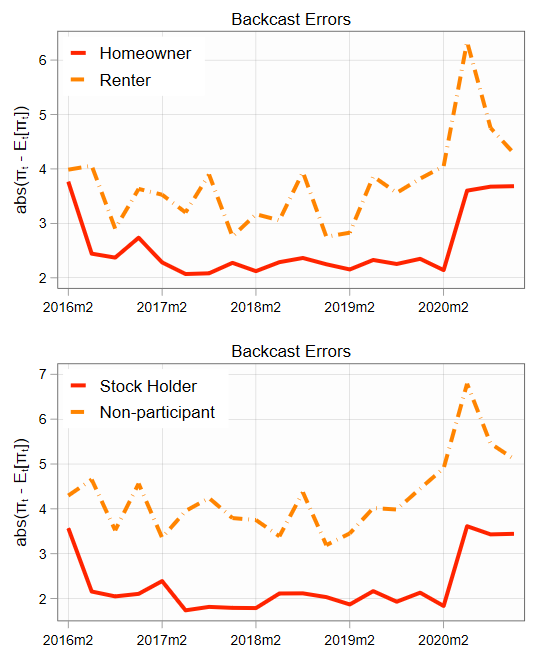} 
    \caption{Backcast accuracy by asset-holding status} 
    \label{fig:error_backcasts}
    
	\noindent\resizebox{1.0\textwidth}{!}{
		\begin{tabularx}{\textwidth}{m{3.8cm}}
			\noalign{\vskip0.3cm}
			\multicolumn{1}{p{\dimexpr\linewidth-2\tabcolsep-2\arrayrulewidth}}{
				\footnotesize \textit{Note:} The average absolute backast errors are reported.} \\
			\multicolumn{1}{p{\dimexpr\linewidth-2\tabcolsep-2\arrayrulewidth}}{	
				\footnotesize \textit{Source:} Authors' calculation. 
            }
		\end{tabularx}
	}
\end{figure}

\begin{figure}[ht!]
    \centering
    \includegraphics[width = 0.55\textwidth]{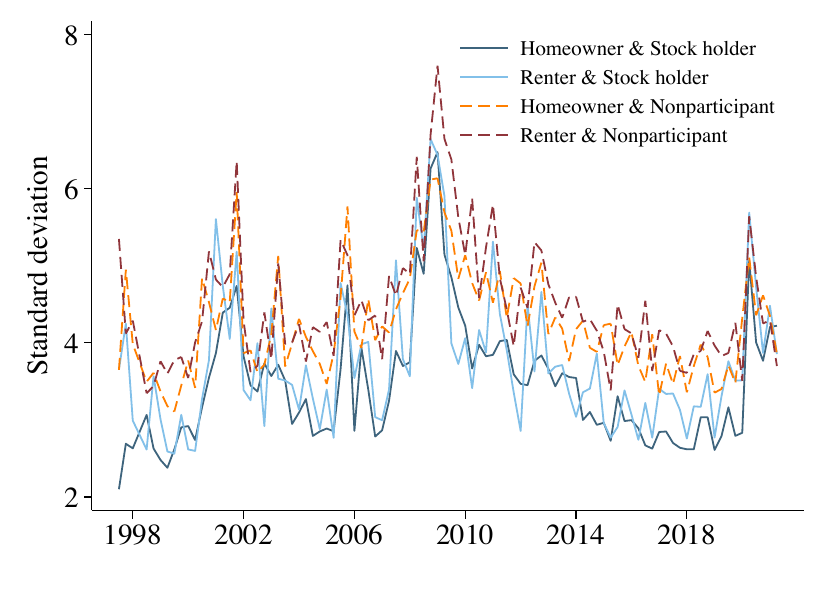}
    \includegraphics[width = 0.55\textwidth]{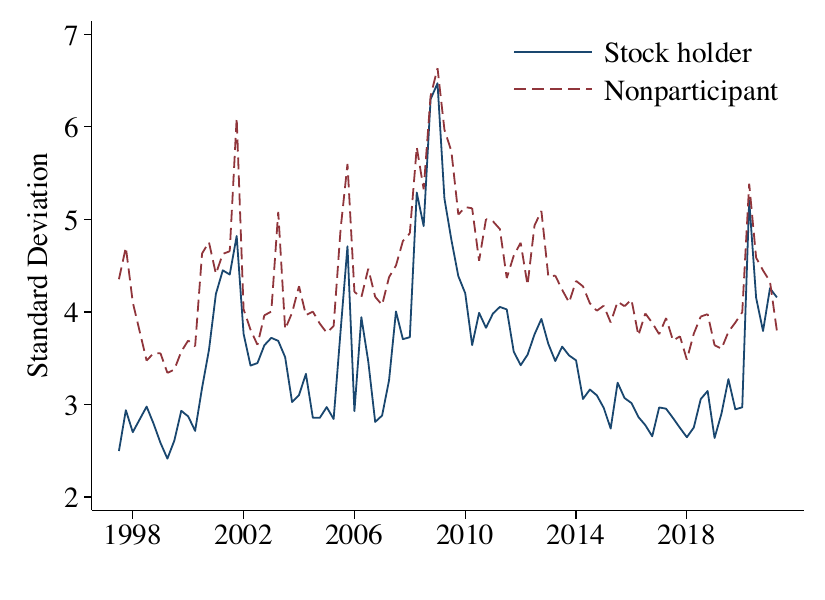}
    \includegraphics[width = 0.55\textwidth]{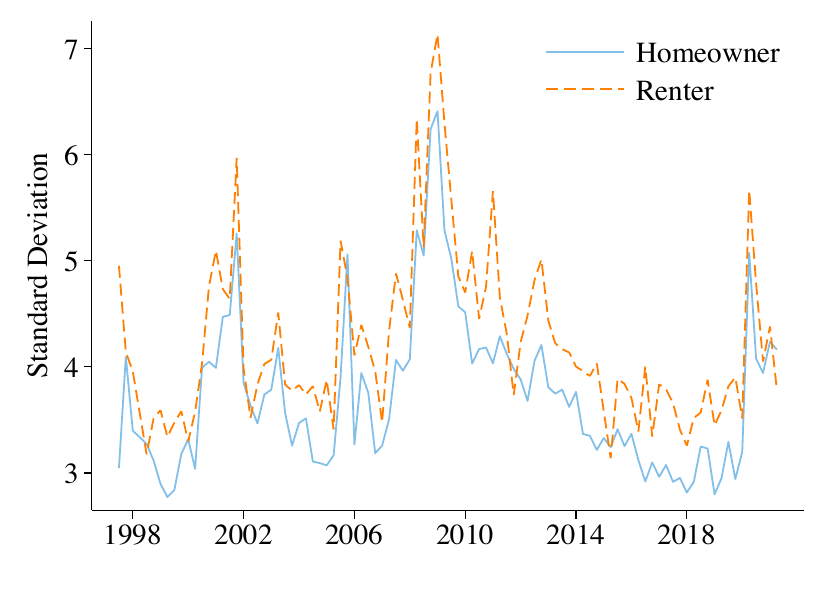}
    \caption{Disagreement stock market participants and non-participants (forecasts)} 
    \label{fig:sd}
    
	\noindent\resizebox{1.0\textwidth}{!}{
		\begin{tabularx}{\textwidth}{m{3.8cm}}
			\noalign{\vskip0.3cm}
			\multicolumn{1}{p{\dimexpr\linewidth-2\tabcolsep-2\arrayrulewidth}}{
				\footnotesize \textit{Note:} The average standard deviation of each group is reported.} \\
			\multicolumn{1}{p{\dimexpr\linewidth-2\tabcolsep-2\arrayrulewidth}}{	
				\footnotesize \textit{Source:} Authors' calculation. 
            }
		\end{tabularx}
	}
\end{figure}

\begin{figure}[ht!]
    \centering
    \includegraphics[width = 0.55\textwidth, height = 0.45\textwidth]{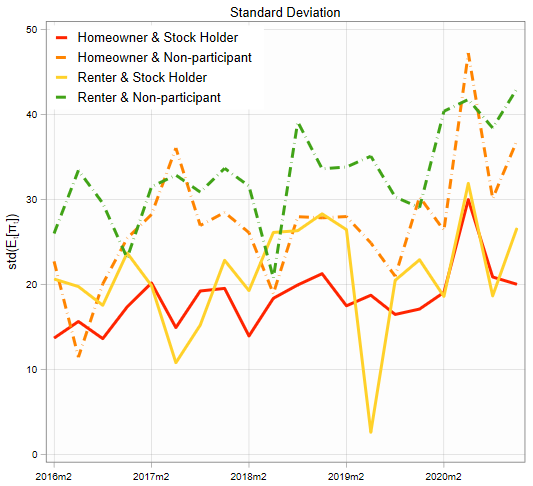} \\ 
    \includegraphics[width = 0.55\textwidth, height = 0.7\textwidth]{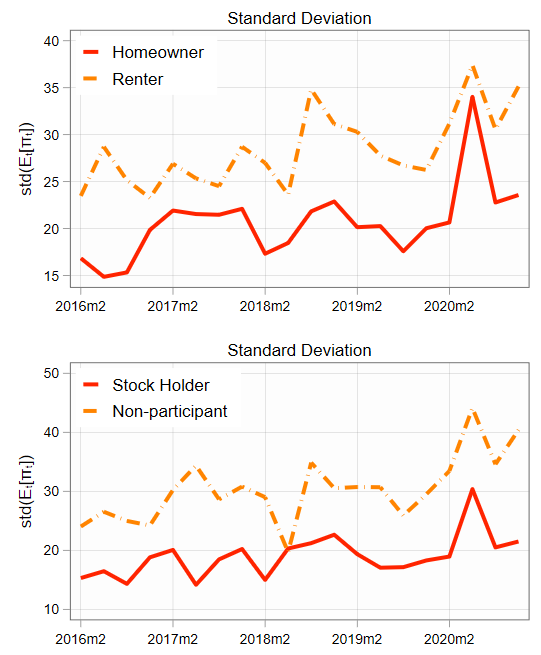} 
    \caption{Disagreement stock market participants and non-participants (backcasts)} 
    \label{fig:sd_backcasts}
    
	\noindent\resizebox{1.0\textwidth}{!}{
		\begin{tabularx}{\textwidth}{m{3.8cm}}
			\noalign{\vskip0.3cm}
			\multicolumn{1}{p{\dimexpr\linewidth-2\tabcolsep-2\arrayrulewidth}}{
				\footnotesize \textit{Note:} The average standard deviation of each group is reported.} \\
			\multicolumn{1}{p{\dimexpr\linewidth-2\tabcolsep-2\arrayrulewidth}}{	
				\footnotesize \textit{Source:} Authors' calculation. 
            }
		\end{tabularx}
	}
\end{figure}


\begin{figure}[ht!]
    \centering
    \caption{Impulse responses of inflation expectations to supply and demand shocks}
    \includegraphics[width = 0.45\textwidth]{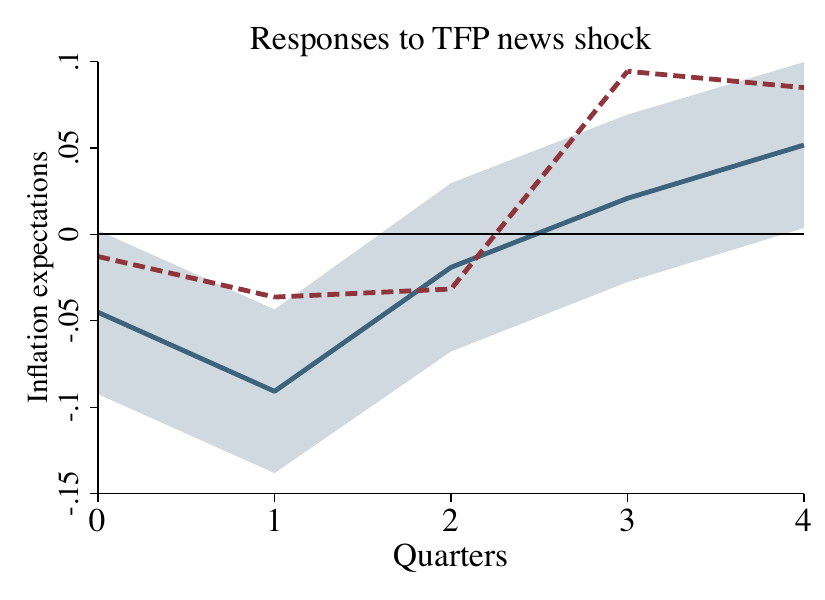}
    \includegraphics[width = 0.45\textwidth]{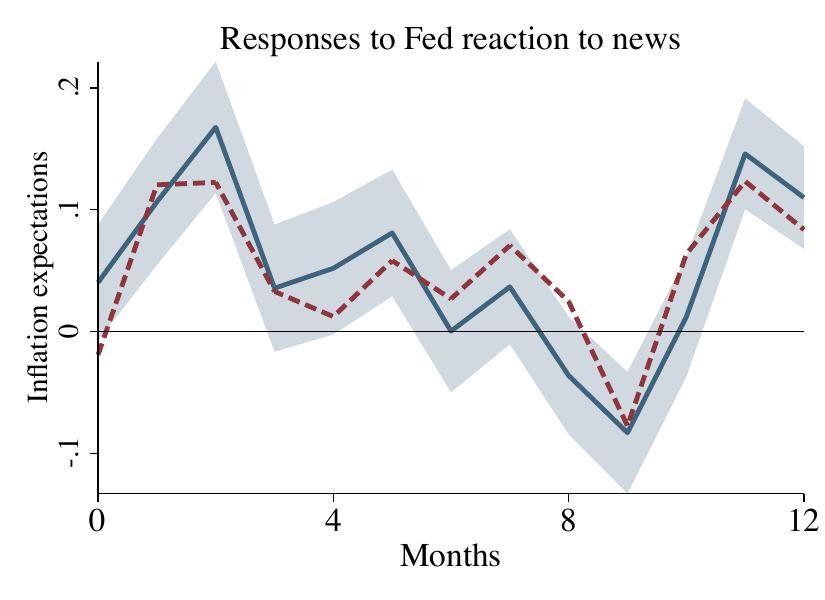}
    \label{fig:irf_inflation}

    \noindent\resizebox{1.0\textwidth}{!}{
		\begin{tabularx}{\textwidth}{m{3.8cm}}
			\noalign{\vskip0.3cm}
			\multicolumn{1}{p{\dimexpr\linewidth-2\tabcolsep-2\arrayrulewidth}}{
				\footnotesize \textit{Notes:} This figure plots the impulse response of stockholders (solid blue line) and non-holders (dashed red line) inflation expectations. The left and right panels are based on TFP news shocks (supply shocks) and Fed reaction to news (demand shocks) respectively. Shaded areas are 95\% CIs.} \\
			\multicolumn{1}{p{\dimexpr\linewidth-2\tabcolsep-2\arrayrulewidth}}{	
				\footnotesize \textit{Source:} Authors' calculation. 
            }
            \end{tabularx}
    }
\end{figure}

\begin{table}[ht!]
    \centering
    \caption{Stock holding and attentiveness: control for attitudes and sentiments} \label{tab:attention_control}
    \small
    \centering
    \vspace{0.3cm}
	\noindent\resizebox{1.0\textwidth}{!}{
	\begin{tabularx}{\textwidth}{ll>{\centering}p{2.7cm}>{\centering}p{2.7cm}>{\centering}p{2.7cm}>{\centering}p{2.7cm}}
        \hline\hline
        \noalign{\vskip0.1cm}
         Dependent variable  & & \multicolumn{2}{c}{\textbf{Forecast errors}} & \multicolumn{2}{c}{\textbf{Attentiveness}}   \tabularnewline 
         &  & (1) & (2) & (3) & (4)  \tabularnewline[0.1cm]
        \hline 
    	\noalign{\vskip0.1cm}
        Stock ($\beta$)  & &  -0.345$^{***}$ &  -0.309$^{***}$  & 0.078$^{***}$ & 0.079$^{***}$ \tabularnewline
                    & &     (0.023)          & (0.023)        &  (0.003)  & (0.003)        \tabularnewline[0.1cm]
        Attitude    & & -0.344$^{***}$ & &  -0.119$^{***}$ \tabularnewline
                    & &  (0.012)        & &  (0.002) \tabularnewline[0.1cm]
        Sentiment    & &        & -0.011$^{***}$  
        & & -0.001$^{***}$ \tabularnewline
                    & &         &  (0.000) 
                    & & (0.000) \tabularnewline[0.1cm]
                    \hline 
        \noalign{\vskip0.1cm}
        Time FE          & & Y & Y & Y & Y  \tabularnewline[0.1cm]
        Demographics FE  & & Y & Y & Y & Y      \tabularnewline[0.1cm]
        Number of obs.  & & 108,026  &  108,026   
        & 120,147  &  120,147      \tabularnewline[0.1cm]
        $R^2$           & & 0.1255   &  0.1373   
        &  0.1305 & 0.1046   \tabularnewline[0.1cm]
        \hline\hline
        
    	\noalign{\vskip0.1cm}
    	\multicolumn{6}{p{\dimexpr\linewidth-2\tabcolsep-2\arrayrulewidth}}{\footnotesize \textit{Notes:} This table reports the regression results from Equations (\ref{eqn:error}) while controlling for attitudes and sentiments. Dependent variables are inflation forecast errors (Columns (1)-(2)) and attentiveness (Columns (3)-(4)) respectively. ``Stock'' indicates dummies for stock market participation. We control for time-fixed effects and survey respondents' demographic fixed effects, including gender, education, birth cohort, marriage status, region, homeownership, and income quintiles. Robust standard errors are reported in the parenthesis.  $^{***}$, $^{**}$, $^{*}$ denotes statistical significance at 1\%, 5\%, and 10\% levels respectively.}    \\ 
    	\multicolumn{6}{p{\dimexpr\linewidth-2\tabcolsep-2\arrayrulewidth}}{\footnotesize \textit{Sources:} Authors' calculation.} 
	\end{tabularx}
    }
\end{table}


\end{document}